\documentclass[
reprint,
superscriptaddress,
amsmath,amssymb,
aps,twocolumn,
prstab,
]{revtex4-2}

\usepackage{graphicx}
\usepackage{dcolumn}
\usepackage{bm}
\usepackage{graphicx}

\usepackage{booktabs}
\usepackage{multirow}
\usepackage{threeparttable}
\usepackage[T1]{fontenc}
\begin{document}


\title{Application of autoresonance in rapid beam extraction of synchrotrons}

\author{X. Ding}
\affiliation{Institute of Modern Physics, Chinese Academy of Sciences, Lanzhou 730000, China}
\affiliation{University of Chinese Academy of Sciences, Beijing 100049, China}
\author{S. Ruan}
\affiliation{Institute of Modern Physics, Chinese Academy of Sciences, Lanzhou 730000, China}
\affiliation{University of Chinese Academy of Sciences, Beijing 100049, China}
\author{H. Ren}
\affiliation{Institute of Modern Physics, Chinese Academy of Sciences, Lanzhou 730000, China}
\author{G. Wang}
\affiliation{Institute of Modern Physics, Chinese Academy of Sciences, Lanzhou 730000, China}
\author{R. H. Zhu}
\affiliation{Institute of Modern Physics, Chinese Academy of Sciences, Lanzhou 730000, China}
\affiliation{University of Chinese Academy of Sciences, Beijing 100049, China}

\author{J. C. Yang}
 \email{yangjch@impcas.ac.cn}
\author{H. Zhao}
 \email{hezhao@impcas.ac.cn}
 
\affiliation{Institute of Modern Physics, Chinese Academy of Sciences, Lanzhou 730000, China}
\affiliation{University of Chinese Academy of Sciences, Beijing 100049, China}

\date{\today}


\begin{abstract} 

In recent years, ultra-high dose rate (FLASH) radiotherapy has become a novel cancer treatment technique because of its similar tumor-killing efficacy as conventional particle therapy while significantly protecting normal tissues. However, due to the limitation of particle number, achieving FLASH condition in a compact heavy-ion synchrotron requires a short extraction time of tens of milliseconds, which is challenging for the conventional RF-KO method. To tackle this challenge, we introduce autoresonance into the third-order resonant extraction for the first time, offering an alternative to the conventional approach of merely increasing the excitation strength. By leveraging a strong detuning effect, a frequency sweeping excitation with small amplitude can drive the entire beam into the autoresonant state, thus enabling rapid beam extraction within a single sweeping period. Compared with the conventional method, this innovative method requires only the addition of an octupole magnet. At the same time, it shows that the conventional RF-KO method has a high autoresonance threshold, so that only a small number of particles that meet the threshold can be excited to large amplitude and be extracted in each sweeping period. In this paper, the autoresonance threshold of a particle in the presence of sextupole and octupole magnetic fields is analyzed, and the single particle simulation shows good agreement with the theoretical formula. Furthermore, the autoresonance based rapid extraction process is simulated and studied, revealing the possibility of millisecond scale beam extraction.

\end{abstract}

\maketitle


\section{introduction}
Particle accelerators play an important role in fundamental and applied research in the sciences, as well as in many technical and industrial fields. Among them, particle therapy has undergone impressive development in the past few decades, due to its excellent physical dose localization at the Bragg peak and a high biological effect, especially for heavy ion beams \cite{tinganelli_ultra-high_2022}. Since the first dedicated therapy facility was built in the 1990s, more than 150 particle therapy facilities are in operation and under construction as of 2024, and more than 410000 patients have been treated worldwide \cite{noauthor_ptcog_nodate}. In recent years, it has been demonstrated that ultra-high dose rates (>40 Gy/s) provide significant normal cell sparing effects while maintaining an equivalent tumor response to conventional dose rates (0.001–0.4 Gy/s) \cite{chow_mechanisms_2024, maxim_flash_2019}. This FLASH effect has great promise and is considered one of the most important discoveries in recent radiotherapy history. 

FLASH irradiation with proton beam are relatively straightforward due to the high beam current from cyclotrons. However, it is challenging to achieve FLASH conditions on carbon ion therapy facilities, which usually use compact synchrotrons and provide much lower dose rates than cyclotrons. Considering the limitation of particle number in compact synchrotrons, a much shorter extraction duration is the most effective way to realize the ultra-high dose rates. Typically, the spill time for FLASH therapy requires reducing proton/ion extraction time in synchrotron from seconds to less than 100 milliseconds, which is the so-called rapid extraction \cite{jolly_technical_2020, colangelo_importance_2019}. Similarly, rapid extraction can also be applied to Rapid Cycling Synchrotrons (RCS) that operate at a repetition frequency above 10 Hz. For example, a parasitic beam slow extraction by foil scattering is proposed in CSNS, which can slowly extract weak proton beam from its 25 Hz RCS while keeping the normal 
fast extraction \cite{zou_parasitic_2014}. 

RF-knockout (RF-KO) method with frequency modulation and amplitude modulation is widely used in synchrotrons for resonance excitation to achieve beam extraction, in which the modulation is mainly to cover the incoherent tunes due to various nonlinear effects \cite{badano1999proton, noda_source_2002, ruan_300_nodate}. Usually, the spill time of conventional RF-KO slow extraction is in the secondary scale, corresponding to the multiple excitation modulation cycles \cite{noda_advanced_2002}. One way to speed up the extraction time to one or several modulation cycles is to increase the strength of the excitation. For example, the Heidelberg Ion-Beam Therapy Center (HIT) achieved carbon beam extraction within approximately 150 ms by increasing the RF driver power to 50 times larger \cite{tinganelli_ultra-high_2022}. However, the high-voltage RF driver increases construction costs and introduces technical challenges, may resulting in certain limitations in its practical application.

In this article, we propose to introduce autoresonance into the RF-KO method, which will effectively reduce the power requirement of the driver and thus efficiently achieve rapid extraction in synchrotrons. Autoresonance is the natural tendency of a weakly driven nonlinear system to remain in resonance with its driver under certain conditions, even if the system parameters vary in time and/or space. It has been well studied in various dynamical systems such as in atomic and molecular physics, plasma physics, fluids and magnetic \cite{fajans_autoresonant_1999-1,fajans_autoresonant_1999,fajans_autoresonant_2001}. Recently, a comprehensive study has also been carried out for the first time in accelerator physics \cite{song_autoresonant_2022}. The autoresonance technique enables nonlinear oscillators to achieve large oscillation amplitudes by applying a swept frequency excitation, in which the beam decoherence effect is minimized and phase-locking is maintained with the drive as long as the excitation amplitude exceeds a specific threshold. Theoretically, this threshold mainly depends on the detuning parameter of the machine and the frequency sweep rate of the driver. The conventional RF-KO extraction method has a large threshold due to the small detuning parameter, making it difficult to achieve the autoresonance condition and cannot excite all particles to a large amplitude at the same time. In this paper, we propose to reduce the autoresonance threshold by intentionally introducing a strong detuning, and demonstrate that this can be achieved using octupoles while ensuring the third-order resonance extraction process is not significantly affected. A comprehensive study is conducted on the autoresonance threshold in the presence of sextupole and octupole. The beam dynamics of the autoresonance and the resulting rapid extraction process are simulated, and the results are consistent with the analytical model.

This paper is organized as follows: Section II examines the autoresonance threshold theory under the simultaneous effects of sextupole and octupole fields, and presents a single particle simulation based on the slow extraction lattice of the SESRI 300 MeV synchrotron \cite{ruan_300_nodate}. Section III discusses the differences between conventional slow extraction and autoresonance rapid extraction, followed by an analysis of beam parameters influence the autoresonance extraction process. Section IV provides the conclusion.

\section{THEORETICAL ANALYSIS AND SIMULATION}
In a nonlinear oscillator physical systems, the oscillation frequency depends on the oscillation amplitude, which is commonly referred to as detuning. To effectively excite the amplitude of such systems, a proven excitation technique called autoresonance can be used. Autoresonance is a phenomenon in nonlinear oscillators where the system automatically adjusts its amplitude in response to changes in the excitation frequency. Theoretically, the threshold of the autoresonance excitation is related to the detuning parameter of the system and the frequency sweep rate of the driver. \cite{fajans_autoresonant_2001}.

\begin{figure}[hbt]
  \includegraphics[width=8cm]{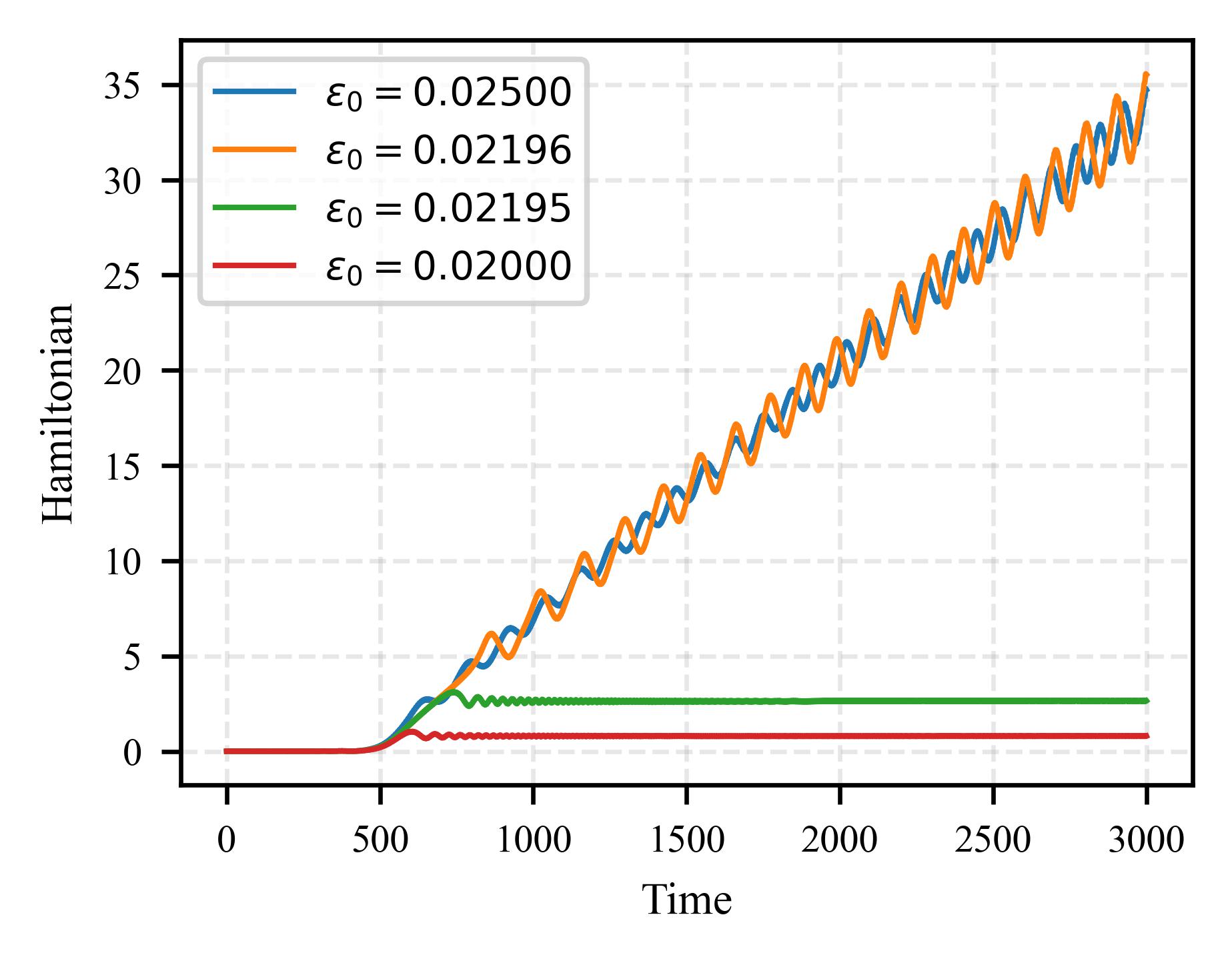}
  \caption{\label{fig:oscillator_response}The response of the oscillator with a nonlinear detuning under different drive strength.}
\end{figure}

For example, we assume a simple oscillator with amplitude $ x = A \cos(\omega_o t + \phi_o) $ oscillating under a excitation amplitude $ \varepsilon = \varepsilon_0 \cos(\omega_d t + \phi_d) $, where the oscillation frequency is linearly related to the system Hamiltonian \( \omega_o = \omega_{o0} + k_1 H \) and \( H = \omega_o^2 x^2 + x'^2/2 \), and the excitation frequency varies linearly with time \( \omega_d = \omega_{d0} + k_2 t \). Fig.~\ref{fig:oscillator_response} shows the response of the oscillator under different drive strengths with a time step of 1 and \( \omega_{o0} = 1.060 \), \( \omega_{d0} = 0.94 \omega_{o0} = 0.996 \), \( k_1 = 0.01 \), and $ k_2 = 1.33 \times 10^{-4} $. As shown in the figure, this oscillator has a very sharp threshold close to \( 0.02196 \). When the excitation amplitude exceeds this threshold, autoresonance has achieved, as illustrated by the orange and blue lines. Throughout the excitation process, the amplitude maintained increases. Conversely, as shown by the red and green lines, the oscillator's amplitude has stabilized at some amplitude, indicating that autoresonance has not been achieved.

In recent years, the ultra-high dose rate (FLASH) radiotherapy has become a novel cancer treatment technique because of its similar tumor-killing efficacy as conventional particle therapy while significantly protecting normal tissues. However, due to the limitation of particle number, achieving FLASH condition in a compact heavy-ion synchrotron requires a short extraction time of tens of milliseconds, which is challenging for conventional RF-KO method. 
The conventional RF-KO method excites particle amplitudes through local resonance, but the extremely high dose rates required for FLASH therapy necessitate large excitation amplitudes that current driver technologies cannot adequately provide within the required millisecond scale. To address this challenge, this paper introduces autoresonance into resonant extraction for the first time. By incorporating a strong detuning effect, this method increases particle oscillation amplitudes in a phase-locked manner using a single-period frequency sweeping excitation.
Leveraging the principles of autoresonance, this technique achieves whole beam amplitude resonance by phase-locked, facilitating rapid and uniform beam extraction. Compared to conventional slow extraction methods, this innovative approach requires only the addition of an octupole magnet to induce a strong detuning effect. Simulation results confirm the theoretical analysis, validating the effectiveness of the proposed Autoresonance Rapid Extraction method.

\subsection{Autoresonance threshold analysis}
In this section, we present an analytical study of the autoresonance threshold in the presence of sextupole and octupole magnetic fields. The Hill's equation of particles in the horizontal direction under an excitation can be simply described by 
\begin{equation}\label{hills}
	{x}^{\prime \prime}+K_1({s}) {x}+\frac{K_2(s)}{2} {x}^2+\frac{K_3(s)}{6} {x}^3=\varepsilon cos \phi_d \sum_n \delta(s-nC) ,
\end{equation}
where $K_1(s)$, $K_2(s)$, $K_3(s)$ represent the field strength of the quadrupole, sextupole and octupole in the longitudinal coordinate $s$, the excitation is regarded as a $\delta$ function per revolution with an amplitude of $\varepsilon$, $n$ is the number of turns and $C$ is the synchrotron circumference. The drive phase $\phi_d = w(n) n + \phi_{d0}$ depends on the linear-sweeping drive frequency $w(n)=w_0 + \alpha n$, where $\alpha$ is the frequency sweep rate. After performing Floquet transformation and canonical transformation, the phase space coordinate $(x,x')$ can be transformed into action-angle space coordinate $(J_x,\Phi_x)$ and the position variable $s$ turns to the azimuth $\theta=2\pi s/C$. The new Hamiltonian for Eq.~(\ref{hills}) expressed in action-angle variables with $\theta$ can be written as
\begin{widetext}
\begin{equation}
	H(J_x,\Phi_x;\theta) = \nu_{x} J_{x} + \frac{1}{2} \alpha_{xx} J_{x}^2 + \frac{\sqrt{2\beta_{x} J_{x}}}{4 \pi} \varepsilon \cos (\Phi_x-\nu_d \theta) + V_3 + V_4,
\end{equation}
\end{widetext}
where $\beta_x$ is the beta function at the location of the excitation, $\nu_d(\theta)=\nu_{d0}+\frac{\theta+2\pi}{4\pi}\alpha = \nu_0+\frac{\alpha}{4\pi}\theta$ is the effective excitation tune. For simplicity, we only consider the first-order nonlinear detuning effect of sextupole and octupole $\alpha_{xx}=\alpha_{xx,s}+\alpha_{xx,o}$. The discrete drive is expanded in a Fourier series, and here we only consider the lowest order term. $V_3$ and $V_4$ represent the third-order and fourth-order resonances excitation terms that caused by sextupole and octupole, respectively. For beam extraction, the betatron tune is close to 1/3, so that the fourth-order resonances can be ignored and the third-order resonance driving term can be approximated by $V_3 = G_{3,0,l} J_x^{3/2} \cos (3 \Phi_x-l \theta+\xi_{3,0,l}) $. The resonance strength $G_{3,0,l}$ can be numerically calculated based on the synchrotron optics \cite{lee2018accelerator}. We assume that the excitation tune sweeps slowly around the extraction tune, i.e. $3\nu_d \approx l$, and perform again the canonical transformation by using the generating function $F_2=\Psi_x J_x = (\Phi_x-\nu_d \theta) J_x$, the new Hamiltonian becomes
\begin{widetext}
\begin{equation}
	\widehat{H}(J_x,\Psi_x;\theta) = (\nu_{x}-\nu_d) J_{x} + \frac{1}{2} \alpha_{xx} J_{x}^2 + k_0 J_{x}^{1/2} \cos \Psi_x + G_{3,0,l} J_{x}^{3/2} \cos3\Psi_x,
\end{equation}
\end{widetext}
where $k_0=\frac{\sqrt{2\beta_{x}}}{4 \pi} \varepsilon$. Now, the particle motion can be described by 
\begin{widetext}
\begin{equation}\label{Eq_motion}
\begin{gathered}
	\frac{d J_x}{d \theta} = k_0 J_x^{1/2} \sin \Psi_x + 3G_{3,0,l} J_{x}^{3/2} \sin3\Psi_x\\
	\frac{d \Psi_x}{d \theta} = \nu_x - (\nu_0 + \frac{\alpha}{2\pi}\theta) + \alpha_{xx} J_x + \frac{1}{2} k_0 J_x^{-1/2} \cos\Psi_x + {\frac{3}{2}}G_{3,0,l} J_{x0}^{1/2} \cos3\Psi_x.
\end{gathered}
\end{equation}
\end{widetext}

Basically, from Eq.~(\ref{Eq_motion}) that when the excitation is off ($\varepsilon=0$) and the third-order resonance strength is small, the particle action will be constant and the phase will advance steadily. When the excitation driver is turns on, the action will increase or decrease depending on the phase, which is also affected by the value of the action $J_{x}$ itself. For a small action, the drive is primarily responsible for the phase changing, and correspondingly resulting in smooth changing of the action. When the action is large, it is difficult for the drive to change the phase and a strong phase mismatch will occur, causing the action stop changing. This is similar to the conclusion about pendulum discussed in Ref.~\cite{fajans_autoresonant_1999,fajans_autoresonant_2001}.

Generally, the phase actually depends on the drive strength $k_0$, frequency sweep rate $\alpha$, detuning parameter of the machine $\alpha_{xx}$ as well as the action value $J_x$. The occurrence of autoresonance exactly means that the system is phase locked and the action response always remain in phase with the drive, so that $\Psi_x$ must remain near a constant. Therefore, we can analyze the threshold for the autoresonance based on the condition of phase lock. We assume $G_{3,0,l}$ is small for particles in the stable region with a small action. Then, the phase $\Psi_x$ can be approximated as being locked around $\pi$ and the action varies smoothly with a small deviation $J_x=J_{x0}+\Delta$, where the equilibrium action $J_{x0}$ can be obtained by 
\begin{equation}\label{eq_action0} 
 \nu_x - (\nu_0 + \frac{\alpha}{2\pi}\theta) + \alpha_{xx} J_{x0} - \frac{k_0}{2} J_{x0}^{-1/2} - {\frac{3}{2}}G_{3,0,l} J_{x0}^{1/2} = 0.
\end{equation}
Substitute Eq.~(\ref{eq_action0}) into Eq.~(\ref{Eq_motion}) and applying the first order Taylor expansion of $J_x^{1/2}$ and $J_x^{-1/2}$, we get 
\begin{equation}
\begin{gathered}
\frac{d \Delta}{d \theta} = k_0 J_{x0}^{1/2} \sin \Psi_x - \frac{\alpha}{2\pi S} + 3G_{3,0,l} J_{x0}^{3/2} \sin3\Psi_x\\
\frac{d \Psi_x}{d \theta} = S \Delta,
\end{gathered}
\end{equation}
where $S=\alpha_{xx}+\frac{1}{4}k_0 J_{x0}^{-3/2} - {\frac{3}{4}}G_{3,0,l} J_{x0}^{-1/2}$. So the Hamiltonian for the new canonical coordinates ($\Delta$,$\Psi_x$) is
\begin{widetext}
\begin{equation}
	\widetilde{H}(\Delta,\Psi_x;\theta) = \frac{1}{2} S \Delta^2 + k_0 J_{x0}^{1/2} \cos \Psi_x + \frac{\alpha}{2\pi S} \Psi_x + G_{3,0,l} J_{x0}^{3/2} \cos3\Psi_x.
\end{equation}
\end{widetext}
Now, the system reduces to a pseudoparticle of slowly varying effective mass $1/S$ moving in a slowly varying pseudopotential $V_{pseudo}= k_0 J_{x0}^{1/2} \cos \Psi_x + \frac{\alpha}{2\pi S} \Psi_x + G_{3,0,l} J_{x0}^{3/2} \cos3\Psi_x$ which is cosine wave with a certain slope. For an autoresonance system that is phase locked, there must be a potential well in which the pseudoparticle is trapped and the phase mismatch is kept close to zero. We still assume the phase locked at $\Psi_x=\pi$, and the condition for the existence of this potential well is simple, i.e.
\begin{equation}\label{eq_final0}
|\frac{\alpha}{2\pi S}|<k_0 J_{x0}^{1/2} + G_{3,0,l} J_{x0}^{3/2}.
\end{equation}
Based on the action of a particle and the setting parameters of the machine, the excitation condition that can drive the particle to autoresonance can be obtained by solving Eq.~(\ref{eq_final0}).

When only octupoles are involved, it can be found that there is a critical action $J_{x0}=(\frac{k_0}{2\alpha_{xx}})^{2/3}$ corresponding to the most difficult situation to achieve autoresonance. Substitute this into Eq.~(\ref{eq_final0}), the drive threshold turns to $\varepsilon > 4 \sqrt{2} \pi \left(\beta \alpha_{xx}\right)^{-\frac{1}{2}} \left(\frac{\alpha}{6 \pi}\right)^{\frac{3}{4}}$ which is applicable to particles with arbitrary actions. With the sweep rate remaining unchanged, the larger the absolute value of $\alpha_{xx}$, the smaller threshold required to achieve autoresonance. This is consistent with the result in Ref.~\cite{song_autoresonant_2022}. Similarly, the autoresonance threshold in the presence of sextupole and octupole can be obtained by using this critical action:
\begin{equation}\label{eq_final1}
-\frac{9 \sqrt{2}}{8} G \sqrt{\beta} \varepsilon+\frac{3}{4(2 \pi)^{1 / 3}} \alpha_{xx}^{\frac{2}{3}} \cdot(\sqrt{\beta} \varepsilon)^{\frac{4}{3}}>\alpha.
\end{equation}

In above, the excitation condition of autoresonance under the influence of sextupole and octupole magnetic fields are studied, which will provide theoretical guidance for the application of autoresonance in rapid beam extraction in synchrotrons. However, some approximations, such as the low-order Fourier approximation of the $\delta$ function and the small third-order resonance strength, may make the theoretical analysis less than perfect. Since we only focus on exciting the beam with small emittance in the stable region to a large amplitude thereby realizing the beam extraction the in the nonlinear region, the assumption that the third-order resonance strength is small in the above derivation is reasonable. And for large-amplitude particles closes to the resonance line, the effect of the sextupole on the resonance strength and the detuning parameter will strongly affect the autoresonance condition, for which the above derivation may not be well applicable. Regardless, a numerical simulation is very necessary for benchmarking with the theory and for more efficient analysis of the autoresonance threshold.

\subsection{Single particle simulation}
\begin{figure}[hbt]
	\includegraphics[width=8.6cm]{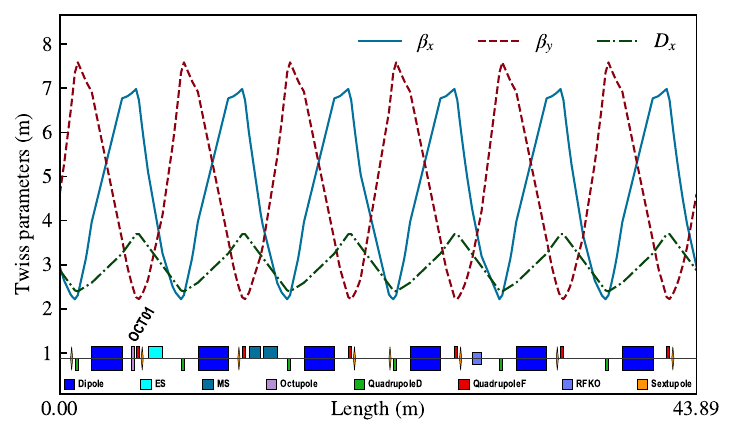}
	\caption{\label{fig:Lattice} Lattice of the SESRI 300 MeV synchrotron.}
\end{figure}

To verify the above conclusion, a single particle simulation is performed based on the slow extraction lattice of SESRI 300 MeV synchrotron at Harbin Space Environment Simulation Research Infrastructure \cite{ruan_300_nodate}. The lattice of the synchrotron is shown in Fig.~\ref{fig:Lattice}, and the major parameters are listed in Table 1. We use the multi-particle tracking code SESP, which is based on linear transfer matrix, and the nonlinear elements are treated as thin-lens approximation. The code has been successfully applied to the design of the SESRI \cite{Gaoyunzhe}. Our simulation employs the 7 MeV/u Bi beam.

\begin{table}[hbt]
\caption{\label{tab-1} Major parameters of the synchrotron}
\renewcommand{\arraystretch}{1.5} 
\begin{tabular}{p{5cm} l c} 
\hline\hline
Circumference $(m)$  &{43.89} \\
Ion species		& p to ${ }^{209} \mathrm{Bi}^{32+}$ \\
Maximal energy $(MeV/u)$	& 300 (p), 7 ( ${ }^{209} \mathrm{Bi}^{32+}$ ) \\
Beam intensity $(ppp)$ & $1\times 10^{9}(\text{p}), 1\times 10^{7}(\text{Bi})$ \\
Magnetic rigidity $( T m)$ & 0.34-2.80 \\
Tune $Q_x / Q_y$	& 1.689/1.736 \\
Natural chromaticity $\xi_x / \xi_y$ & -0.71/-1.03 \\
$3^{rd}$ order resonance strength \newline $G_{3,0,5}$ $(m^{-1/2})$ & 0.00608\\
$1^{st}$ order detuning parameter \newline caused by sextupoles $(m^{-1})$ & -25.99 \\
Octupole beta function $(m)$ & $\beta_x / \beta_y=6.89 / 2.57$ \\
RF-KO beta function $(m)$ & $\beta_x / \beta_y=4.27 / 3.16$ \\
\hline\hline
\end{tabular}
\end{table}

There are eight sextupole magnets symmetrically distributed in the synchrotron. Based on the design value of sextupoles, the theoretical strength of the third-order resonance $G_{3,0,5}$ is about $0.00608~m^{-1/2}$, and the calculated first order nonlinear detuning parameter $\alpha_{xxs}$ is about $-25.99~m^{-1}$. In the simulation, an octupole is introduced to achieve strong detuning thereby reducing the autoresonance threshold. The octupole is located near a horizontal focusing quadrupole, to minimizes the effect of the nonlinear field on the vertical motion. Assuming the octupole length is $l=0.15~m$ and the coefficient is $K_3=-250.5~m^{-4}$, a detuning parameter $\alpha_{xxo} = -47.30~m^{-1}$ is obtained. 

\begin{figure}[hbt]
	\includegraphics[width=8.4cm]{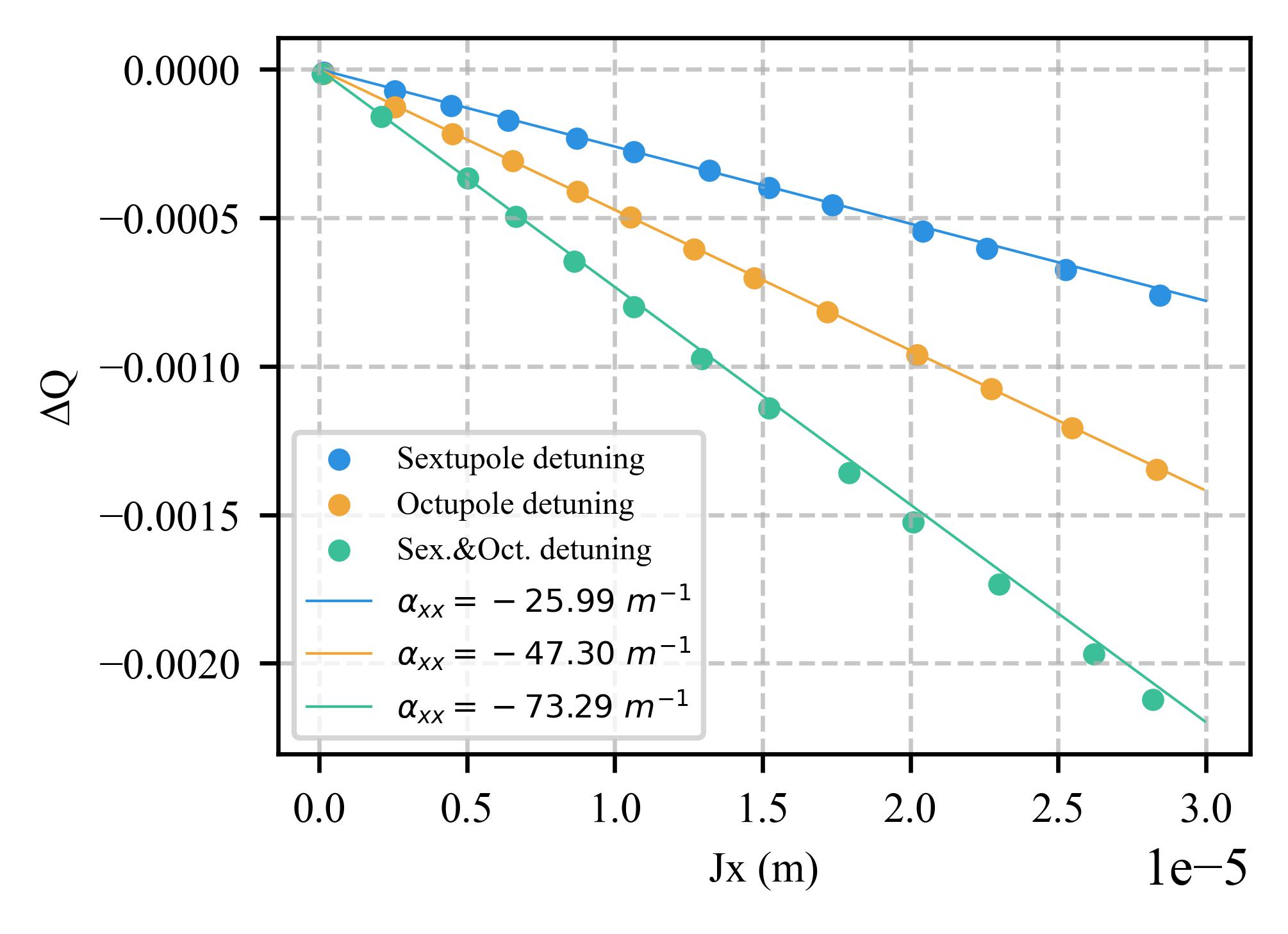}
	\caption{\label{fig:tune_shift}Comparison of simulation and theoretical results for the tune shifts with only sextupole, only octupole and both.}
\end{figure}

The detuning parameters is tracked according to particles betatron motion with different actions, by the PyNAFF \cite{laskar_chaotic_1990, PyNAFF} code . As shown in Fig.~\ref{fig:tune_shift}, the dependence of the tune shift on the action for only sextupoles, only octupoles, and both are obtained. The tracking detuning parameters are consistent with the theoretical values \cite{lee2018accelerator}. However, we do not consider the higher order detuning near the third-order resonance, which could be important for the particles with large actions that close to the extraction orbit.

In the simulation, a cosine wave driver with a unidirectional frequency sweep is employed. The frequency follows $f_d=f_s+(f_e-f_s)\frac {n}{N}$, where $n$ is the current number of turns, and $N$ is the total simulation number of turns which is used to adjust the frequency sweep rate. Start from the base frequency, the excitation frequency sweep down/up depending on the sign of the detuning parameter. In our case, the base frequency is $f_s = 1.4103 ~\text{MHz}$ which we set slightly above $\nu_x f_0$ to cover all the particles with different emittance, where $\nu_x$ is the linear betatron tune and $f_0$ is the repetition frequency. And the final frequency $f_e = 1.3928~ \text{MHz}$ corresponds to the design extraction tune 1.676. 

\begin{figure}[hbt]
  \centering
  \begin{minipage}{0.23\textwidth}
    \centering
    \includegraphics[width=\linewidth]{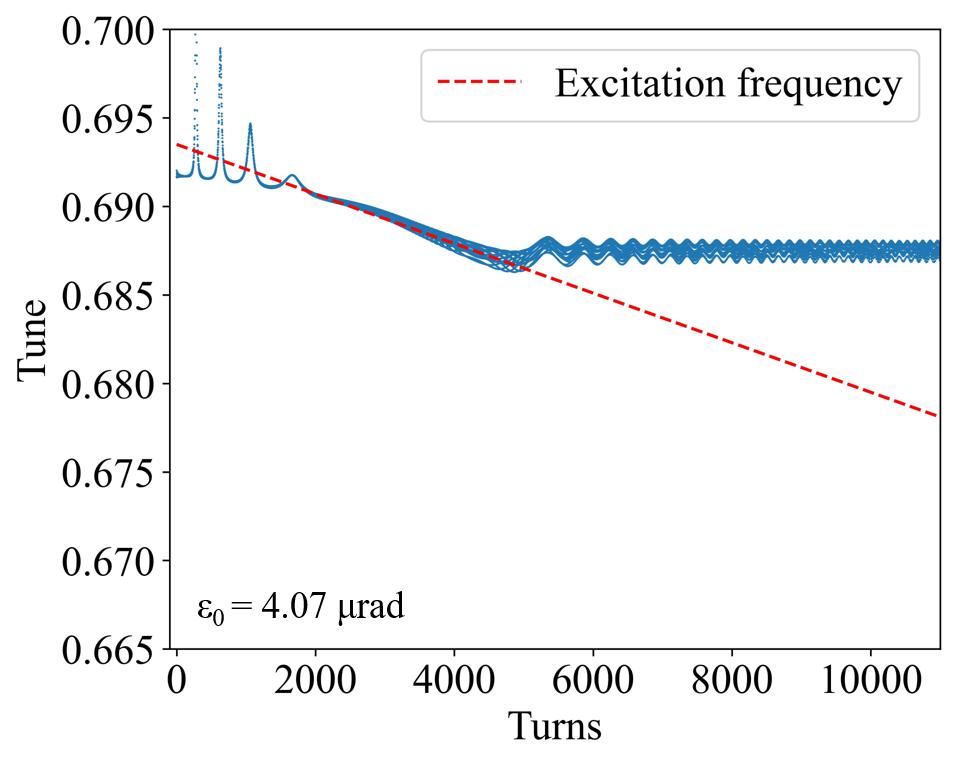}
  \end{minipage}
  \hspace{0\textwidth} 
  \begin{minipage}{0.23\textwidth}
    \centering
    \includegraphics[width=\linewidth]{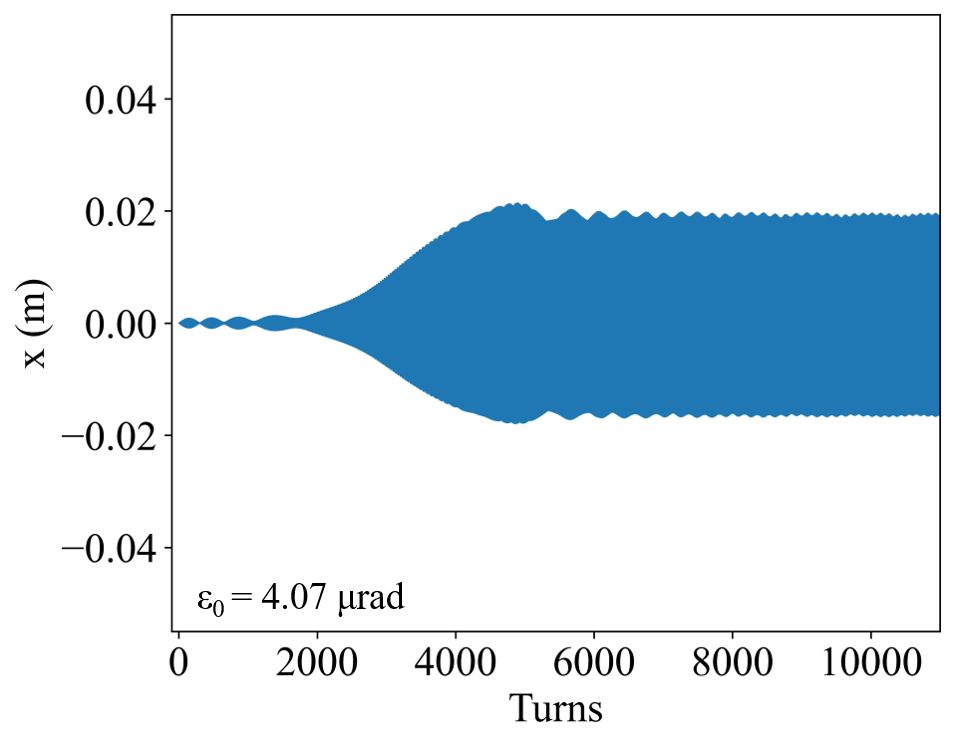}
  \end{minipage}

  \vspace{0.5em} 

  \begin{minipage}{0.23\textwidth}
    \centering
    \includegraphics[width=\linewidth]{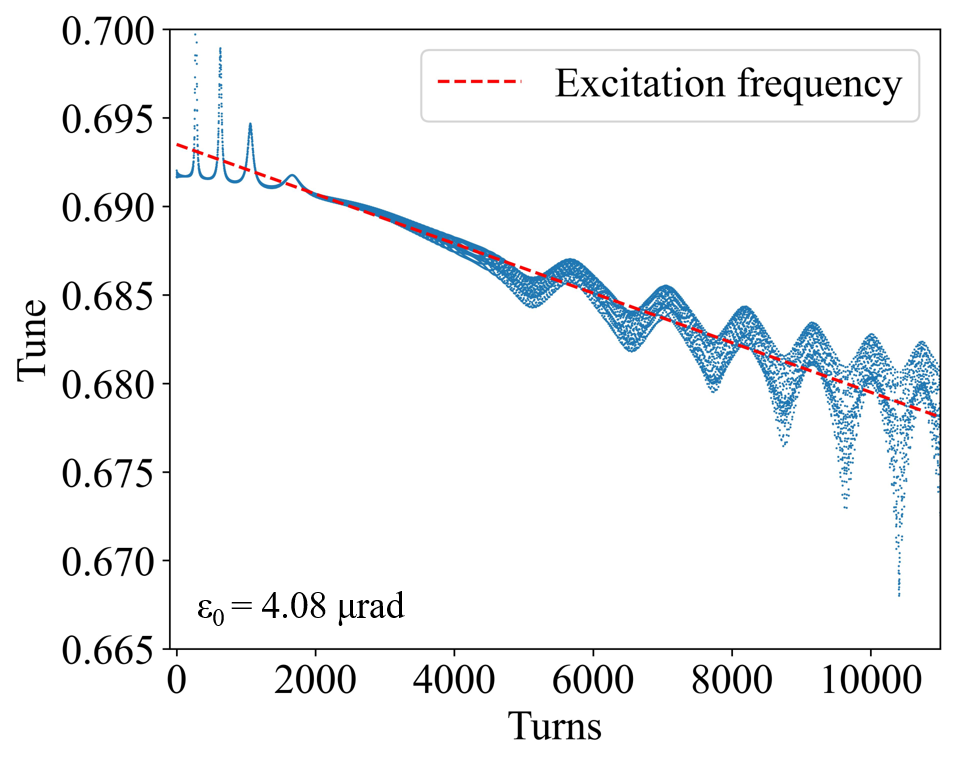}
  \end{minipage}
  \hspace{0\textwidth} 
  \begin{minipage}{0.23\textwidth}
    \centering
    \includegraphics[width=\linewidth]{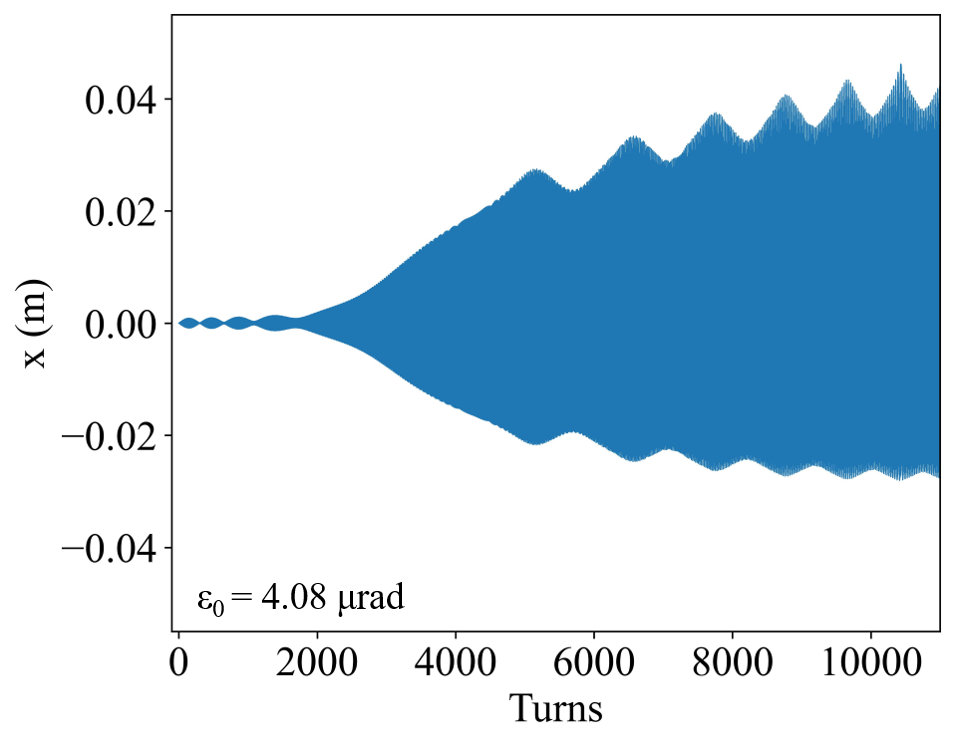}
  \end{minipage}
  
  \caption{\label{fig:occurrence_of_autoresonance} Evolution of the particle (initial emittance of $1\times 10^{-5}~ \mu m$) tune and betatron amplitude below (up) and exceed (down) the threshold which is between $4.07~\mu rad$ and $4.08~\mu rad$.}
\end{figure}

We observe the variation of the betatron amplitude of a particle over the number of turns to determine whether the particle reaches the autoresonance.As shown in Fig.~\ref{fig:occurrence_of_autoresonance}, the evolution of the oscillating tune and betatron amplitude of the particle under different excitation amplitudes are presented. For the excitation amplitude of $\varepsilon_0=4.07~\mu rad$, the drive phase and the particle oscillating phase are mismatched after 5000 turns so autoresonance is not excited. For the excitation amplitude of $\varepsilon_0=4.08~\mu rad$ that exceeds the threshold, it results in a maintained increase of particle amplitude and phase lock between particle oscillation and the drive. It is clear that the autoresonance threshold is quite sharp and there is a significant difference in particle motion exceed and below the threshold. 

\begin{figure}[hbt]
	\includegraphics[width=8.4cm]{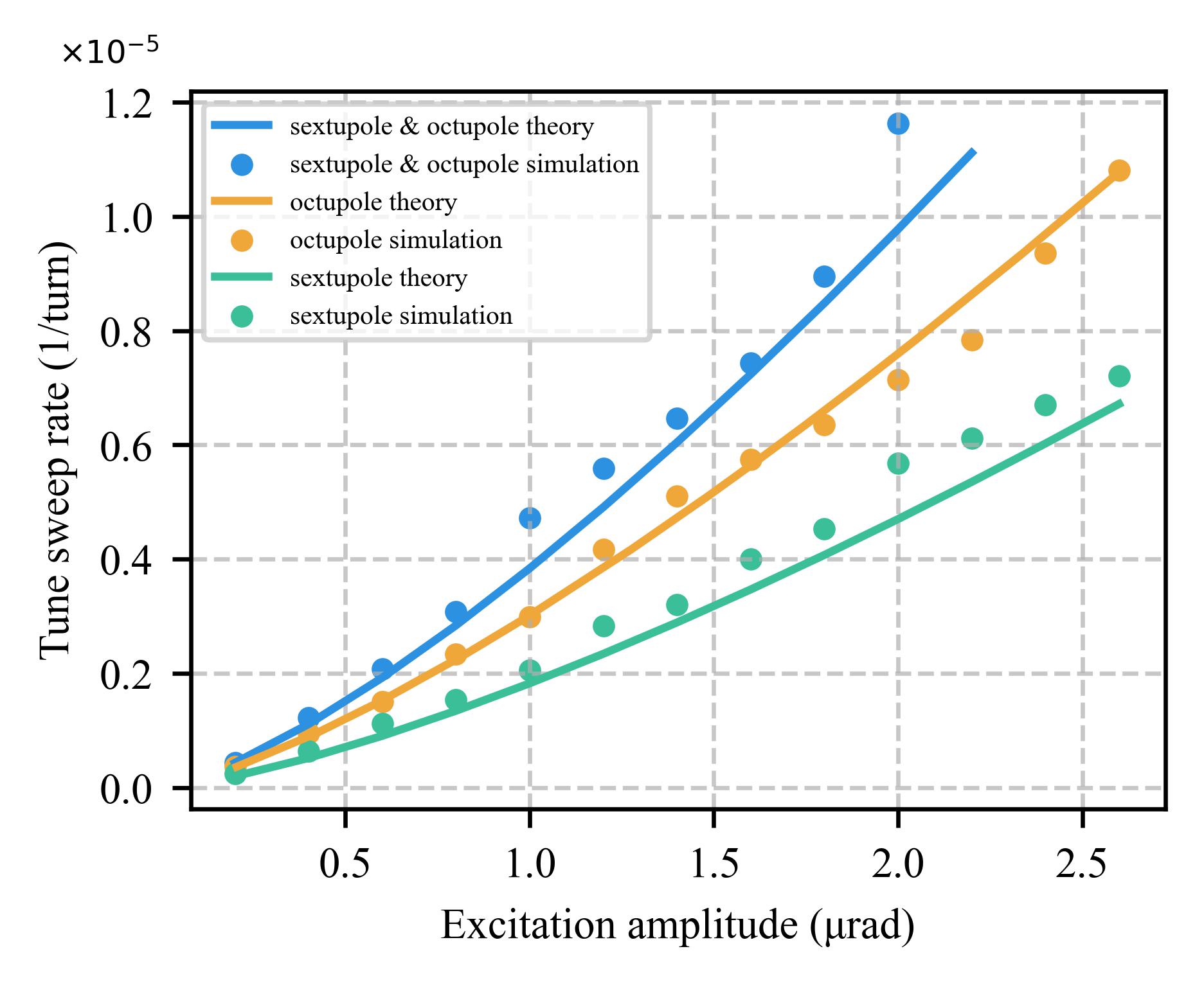}
	\caption{\label{fig:threshold} The relationship between the autoresonance driving threshold on the frequency sweep rate, which shows the simulation results are consistent with the theoretical calculations.}
\end{figure}

Taking whether the phase is maintained locked throughout the entire simulation process as the criterion, the autoresonance threshold under different conditions are calculated. Fig.~\ref{fig:threshold} shows the relationship between the excitation threshold and the frequency sweep rate for the cases with only octupole, only sextupole, and both. The theoretical framework is consistent with those obtained from particle simulations. In particular, the result for the case with only an octupole agrees well with those presented in the Ref.~\cite{song_autoresonant_2022}. However, it is worth noting that the linear betatron tune we set is slightly away from one third, making the third-order resonance strength $G$ small. This is consistent with the assumption introduced in Sec. II, so that the theoretical curves with sextupoles are in good agreement with the simulation. 

It also demonstrates that the extra detuning introduced by the octupole can reduce the excitation threshold, thus achieving a more effective rapid extraction method than the conventional RF-KO slow extraction. Theoretically, the requirement of the excitation can be significantly reduced by a stronger detuning, as long as it will not strongly affect the extraction separatrix. In the subsequent section, the rapid extraction method based on the autoresonance will be introduced, and the differences and advantages over the conventional slow extraction will be discussed.

\section{AUTORESONANCE based RAPID EXTRACTION}

In the conventional RF-KO slow extraction method, the tune needs to be adjusted to be close to the third-order resonance line by quadrupole magnets after acceleration. Single-frequency or dual-frequency excitation with a sweep range of several hundred hertz is applied to drive particle oscillation amplitude \cite{noda_advanced_2002}. When the oscillation amplitude of the particles reaches the phase-stable triangular separatrix formed by the sextupole magnets, the particles follow this separatrix into the extraction channel, resulting in successful extraction. During this process, a small bunch of particles is excited and extracted with each excitation sweep cycle, enabling control over beam spill.

\begin{figure}[hbt]
	\includegraphics[width=8.4cm]{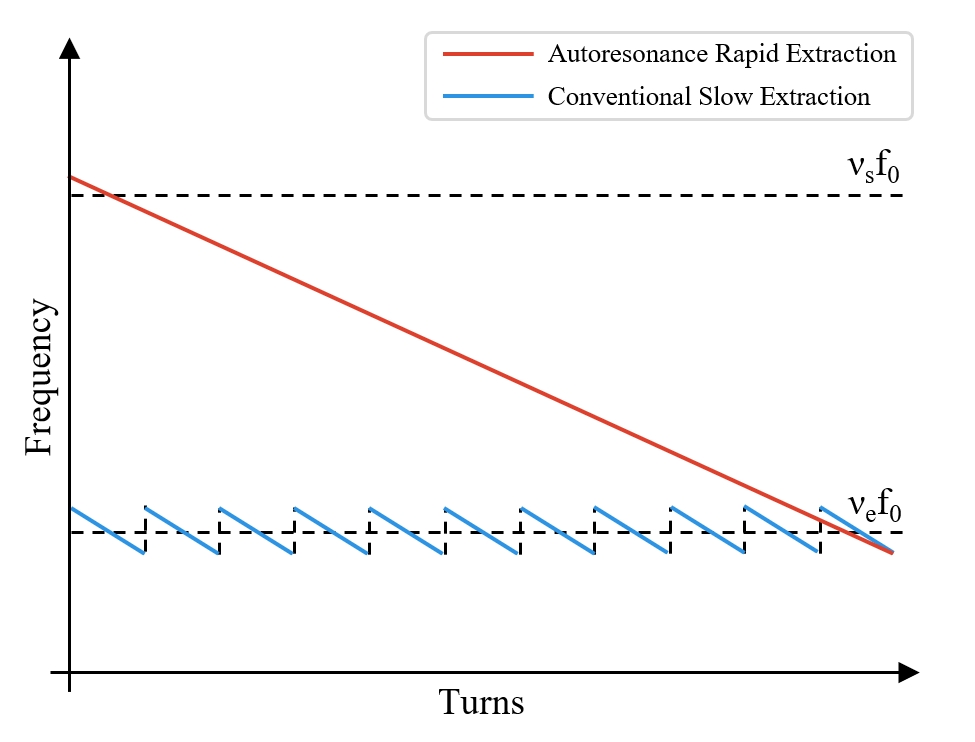}
	\caption{\label{fig:sweep}Comparison of autoresonance rapid extraction frequency sweeping curve (red) and conventional slow extraction single-frequency sweeping curve (blue).}
\end{figure}

\begin{figure}[hbt]
  \centering
  \begin{minipage}{0.24\textwidth}
    \centering
    \includegraphics[width=\linewidth]{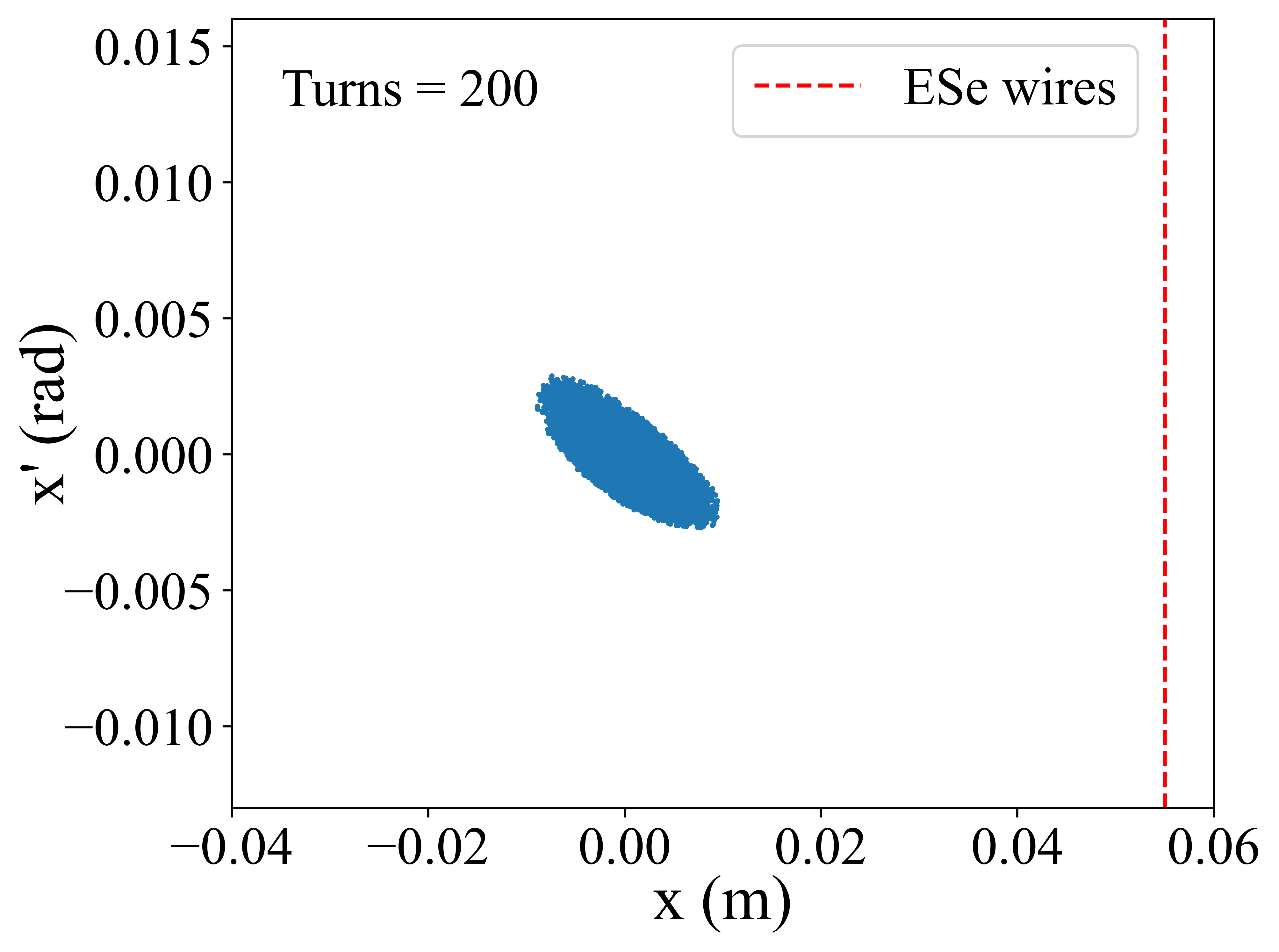}
  \end{minipage}
  \hspace{-0.01\textwidth} 
  \begin{minipage}{0.24\textwidth}
    \centering
    \includegraphics[width=\linewidth]{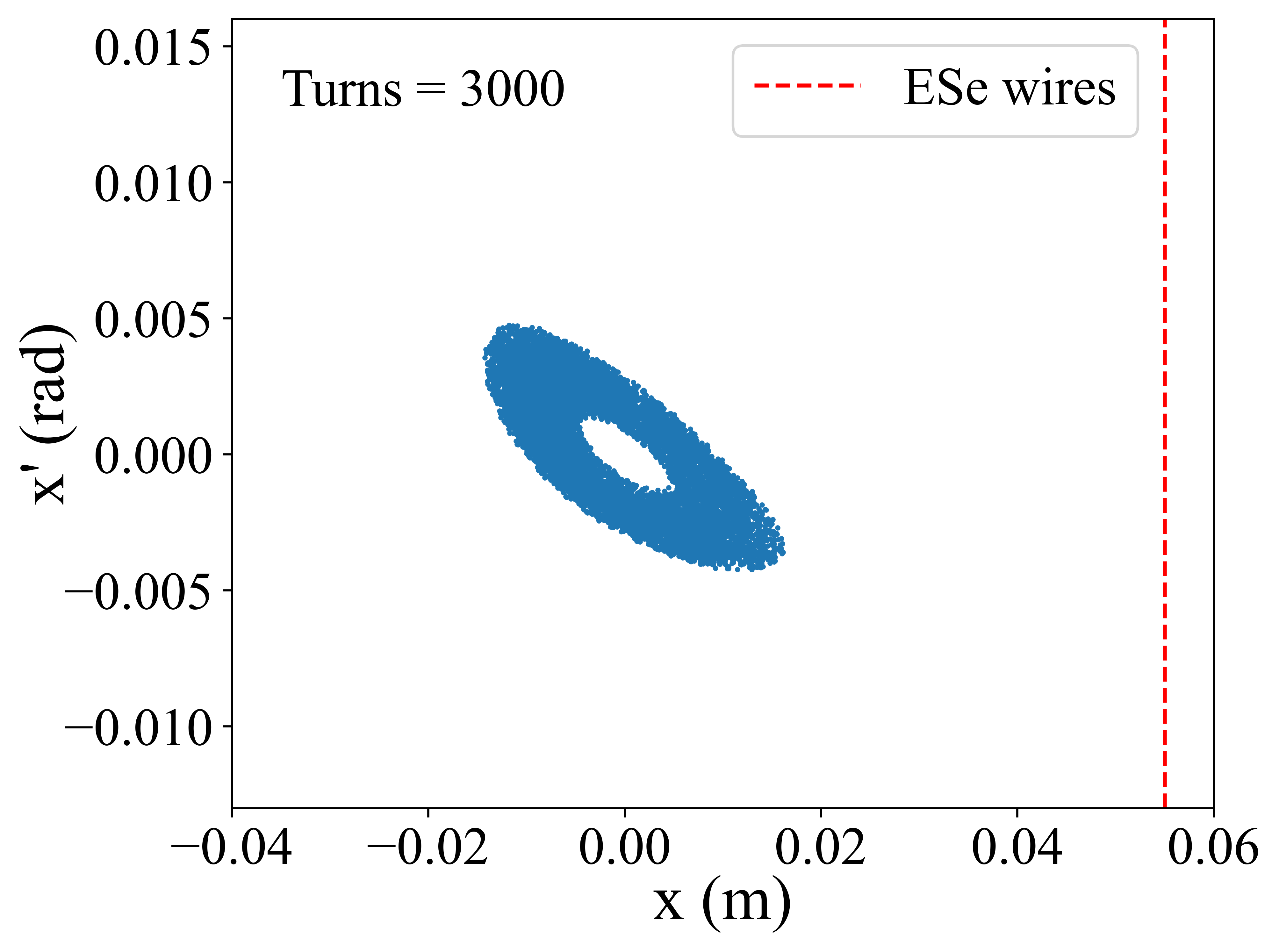}
  \end{minipage}

  \vspace{0.5em} 

  \begin{minipage}{0.24\textwidth}
    \centering
    \includegraphics[width=\linewidth]{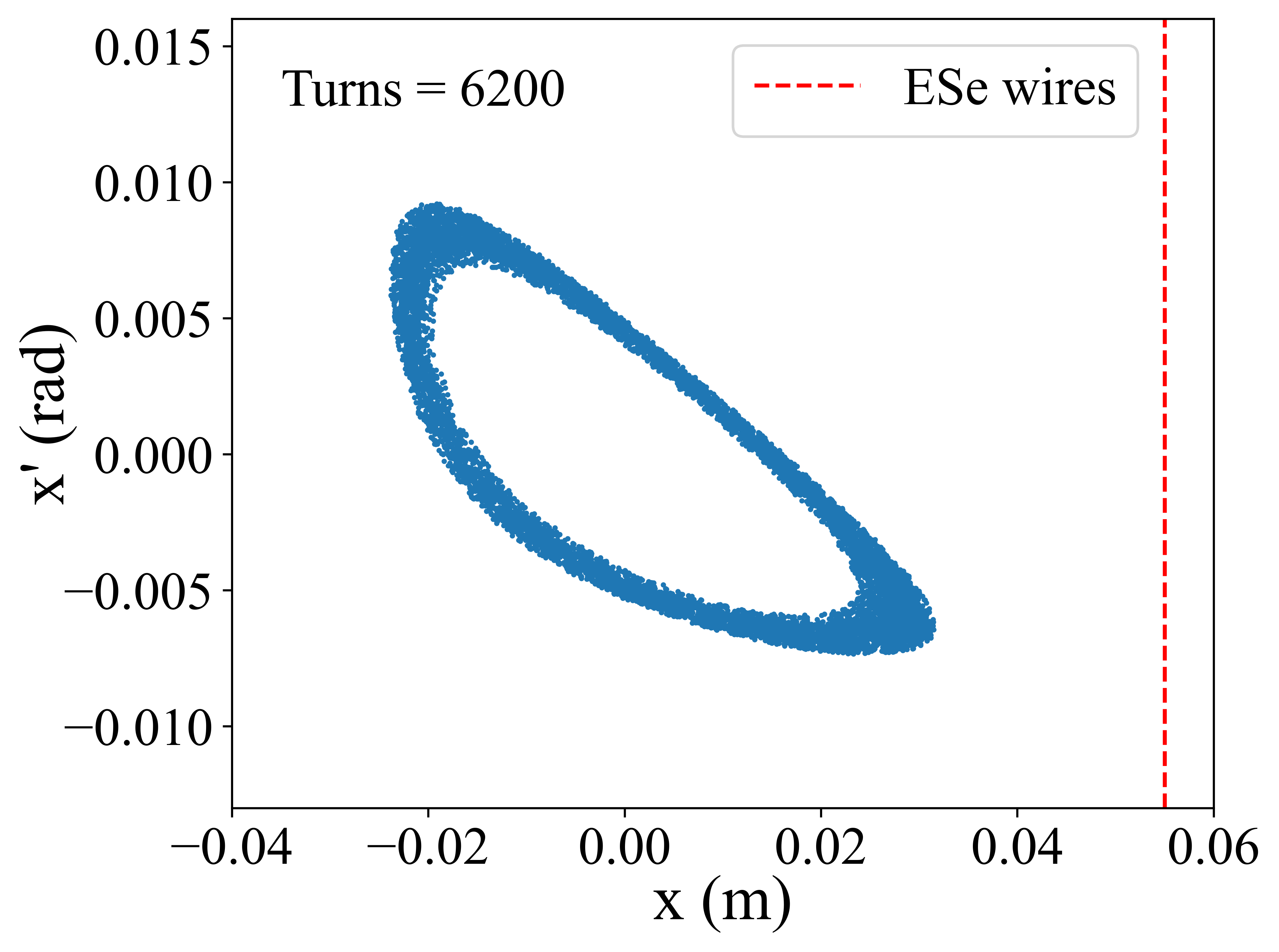}
  \end{minipage}
  \hspace{-0.01\textwidth} 
  \begin{minipage}{0.24\textwidth}
    \centering
    \includegraphics[width=\linewidth]{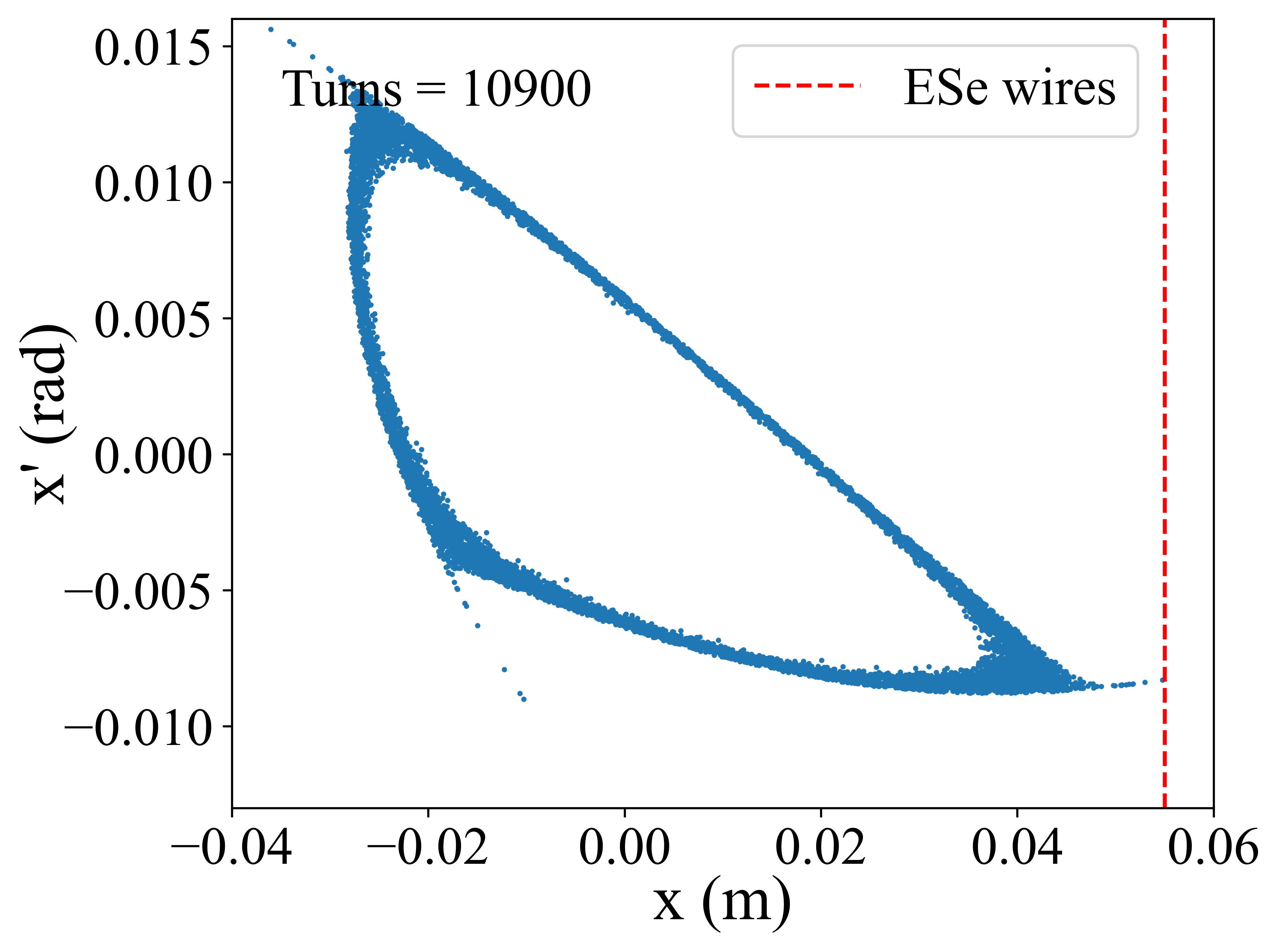}
  \end{minipage}

  \caption{\label{fig:process_of_autoresonant_rapid_extraction} Simulation results of the beam distribution in phase space during the process of the autoresonance based rapid extraction. The excitation amplitude is $10~\mu rad$ and the frequency sweep rate is $1.4\times 10^{-6}$. The location of ESe wires is indicated by the red dashed line.}
\end{figure}

Autoresonance rapid extraction employs single-period frequency sweeping excitation, as illustrated in Fig.~\ref{fig:sweep}, with a sweep range from $\nu_s f_0~to~\nu_e f_0$, where $\nu_s$ is the linear tune and $\nu_e$ is the design extraction tune and $f_0$ is repetition frequency. There is no need for specific adjustment of the accelerator's linear tune in the entire acceleration. As the amplitude increases, the tune of particles automatically converges towards the resonance line due to the detuning effects caused by the octupole and sextupole magnets.  The single-period frequency sweeping excitation causes the oscillation amplitudes of all particles increase collectively, as shown in Fig.~\ref{fig:process_of_autoresonant_rapid_extraction}, and resulting in only one distribution spill structure.

In the conventional RF-KO slow extraction, the excitation amplitude fails to exceed the autoresonance threshold for most particles, so only a portion of particles are excited during each frequency sweep. The initial conditions of the other particles are changed by the previous excitation, as shown in the upper right figure of Fig.~\ref{fig:occurrence_of_autoresonance}. Consequently, in the next frequency sweep period, new particles may exceed the autoresonance threshold and can strongly excited and extracted. Ultimately, the beam forms a seemingly continuous extraction over seconds, but microscopically, it exhibits a spill structure characterized by the frequency sweep period as many Rayleigh distributions. In contrast, the autoresonance rapid extraction method utilizes strong detuning to reduce the autoresonance threshold. During this process, the pseudoparticle remains within the pseudopotential well of the Hamiltonian at all times. As a result, all particles in the beam are strongly excited within a single-period frequency sweeping excitation, reducing the extraction time to the millisecond-scale. The spill structure of this method exhibits a single Rayleigh distribution, but the desired uniform distribution can be achieved through standard techniques such as pre-set excitation amplitude modulation and real-time feedback system. Additionally, the strong detuning requires the initial tune to be set further away from the third-order resonance line, which is benefit for the beam's stability. We find that the two extraction modes differ in mechanism, but their final extraction effects show notable similarities. This represents a new understanding of the slow extraction process, which coincides with the theoretical analysis in this paper and the spill structure simulation results in the next section. It may also provide new insights for future beam extraction technologies. 

\begin{figure}[hbt]
  \centering 
  \includegraphics[width=8.4cm]{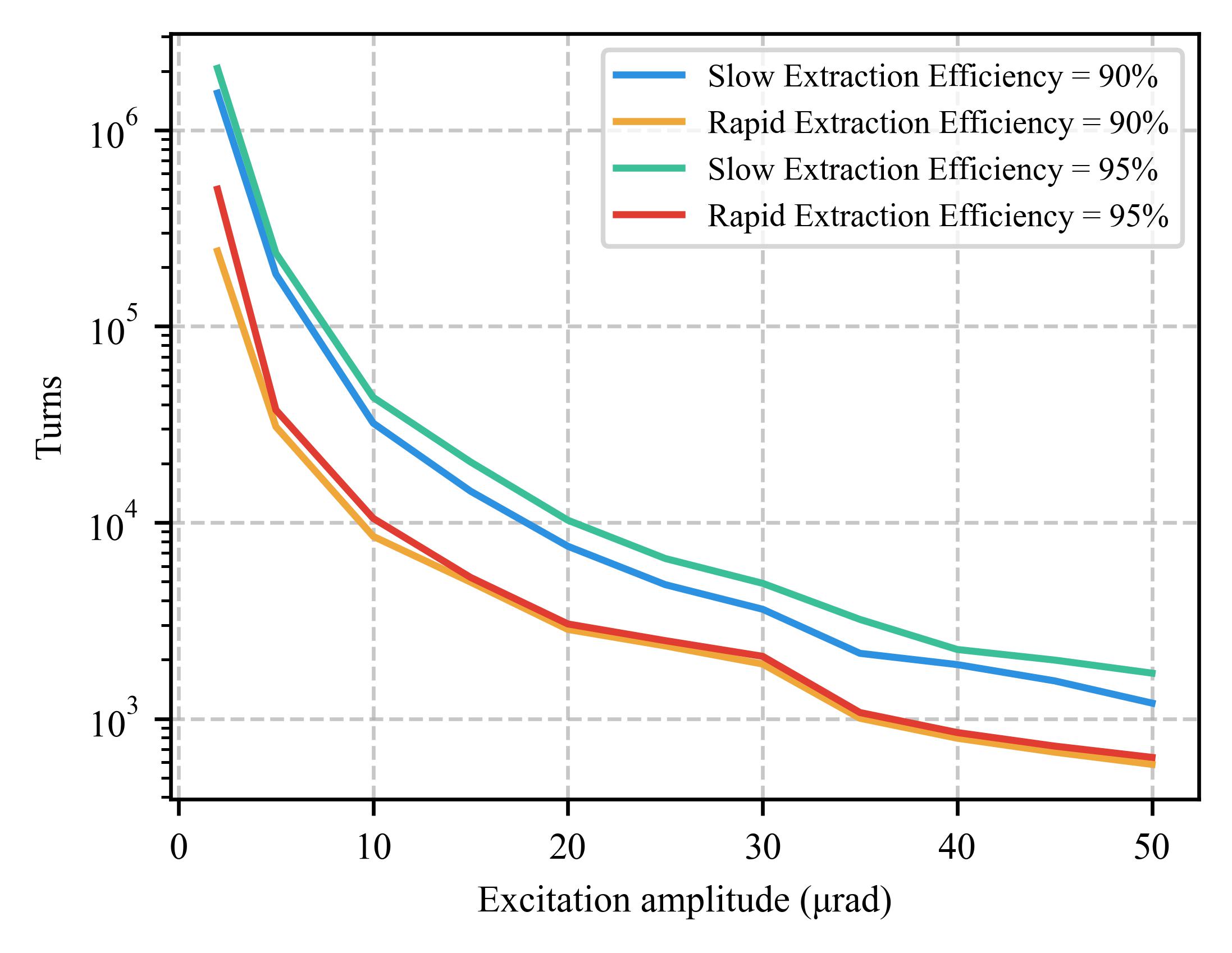} 
  \caption{\label{fig:slow_extraction_and_autoresonance_rapid_extraction} Comparison of the number of turn required for conventional slow extraction and autoresonance rapid extraction when the extraction efficiency reaches 90\% and 95\%, respectively.}
\end{figure}

As shown in Fig.~\ref{fig:slow_extraction_and_autoresonance_rapid_extraction}, to compare the speeds of the two extraction methods, we recorded the minimum number of turns required to achieve the same extraction efficiency (90\%, 95\%) under the same excitation amplitude. The results indicate that autoresonance rapid extraction is nearly an order of magnitude faster than conventional slow extraction at all excitation intensities, the extraction time required for the autoresonance rapid extraction is about several milliseconds. With sufficiently high excitation amplitudes, it can even reach the submillisecond scale. It is much faster than the fastest extraction time achievable by conventional slow extraction. Moreover, conventional slow extraction requires time to adjust the tune, whereas the new extraction method is even faster in application by comparison.

\subsection{The influence of beam initial emittance and momentum spread}

To investigate the effect of the beam emittance on autoresonance excitation, simulations are conducted in the SESRI 300 MeV synchrotron for beams with different initial emittance and zero momentum spread. The excitation amplitude is set to \( 10 \, \mu\mathrm{rad} \), and tracking is performed over 15000 turns.

\begin{figure}[hbt]
  \centering
  \begin{minipage}{0.23\textwidth}
    \centering
    \includegraphics[width=\linewidth]{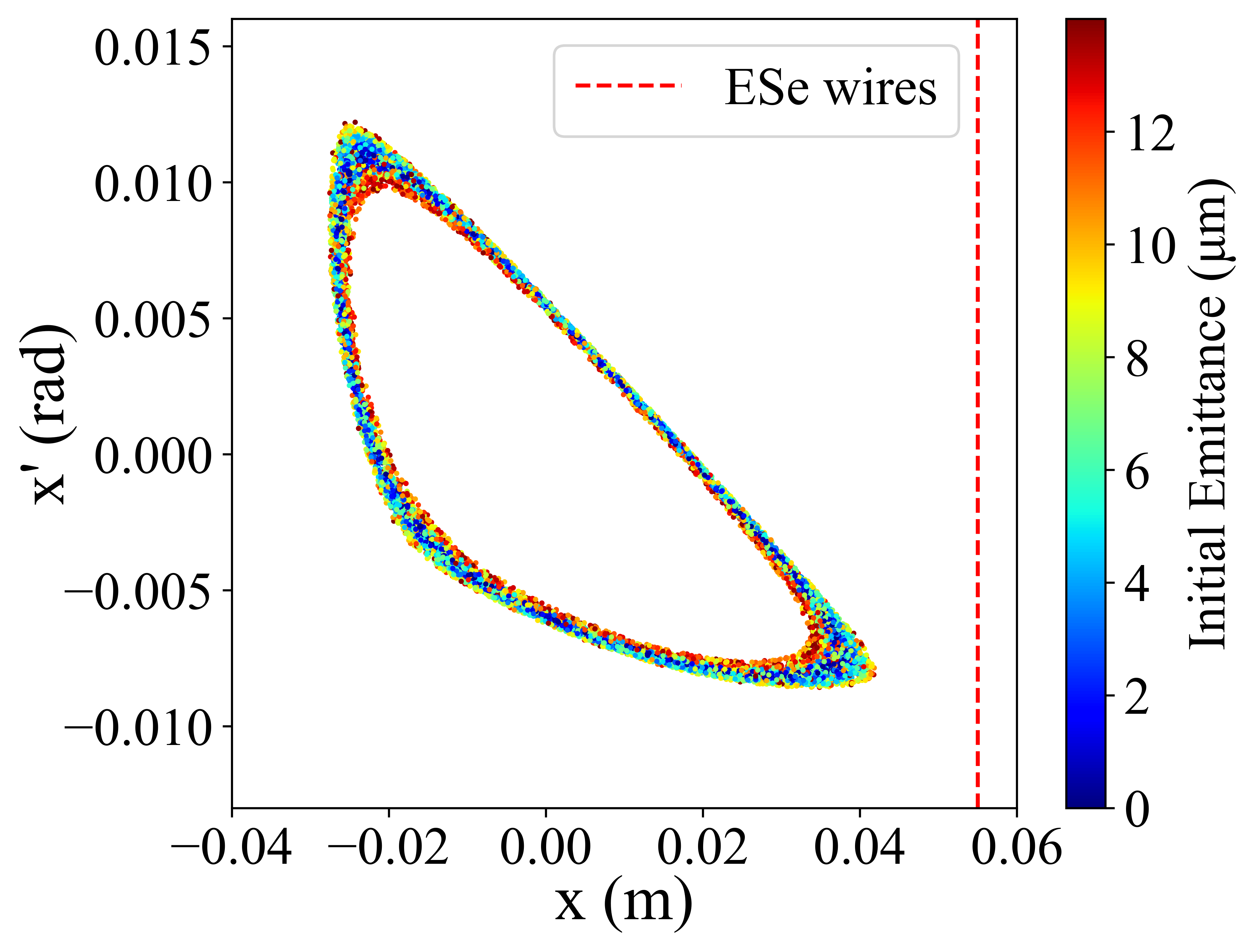}
  \end{minipage}
  \hspace{0.004\textwidth} 
  \begin{minipage}{0.23\textwidth}
    \centering
    \includegraphics[width=\linewidth]{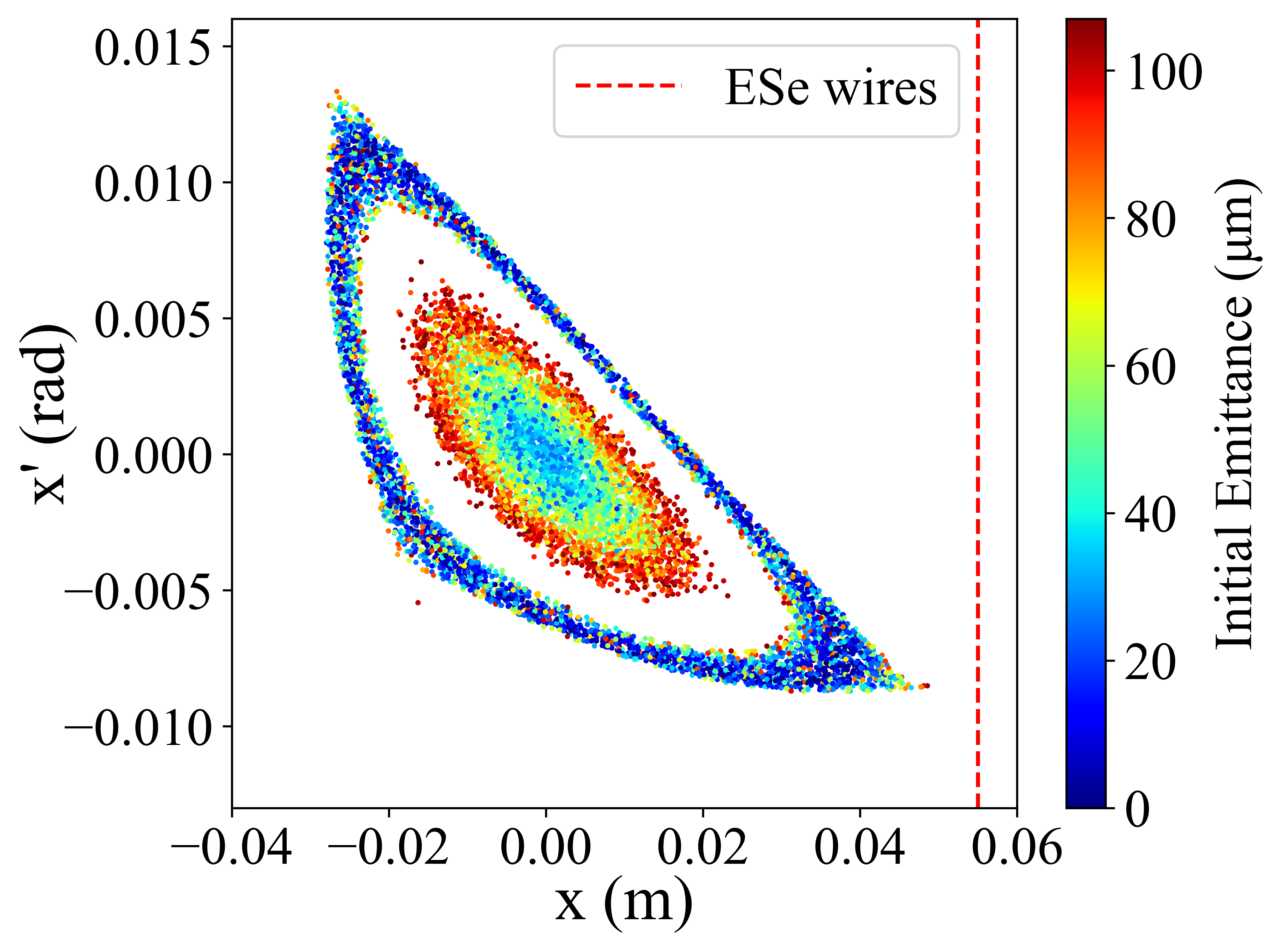}
  \end{minipage}

  \vspace{0.5em} 

  \begin{minipage}{0.23\textwidth}
    \centering
    \includegraphics[width=\linewidth]{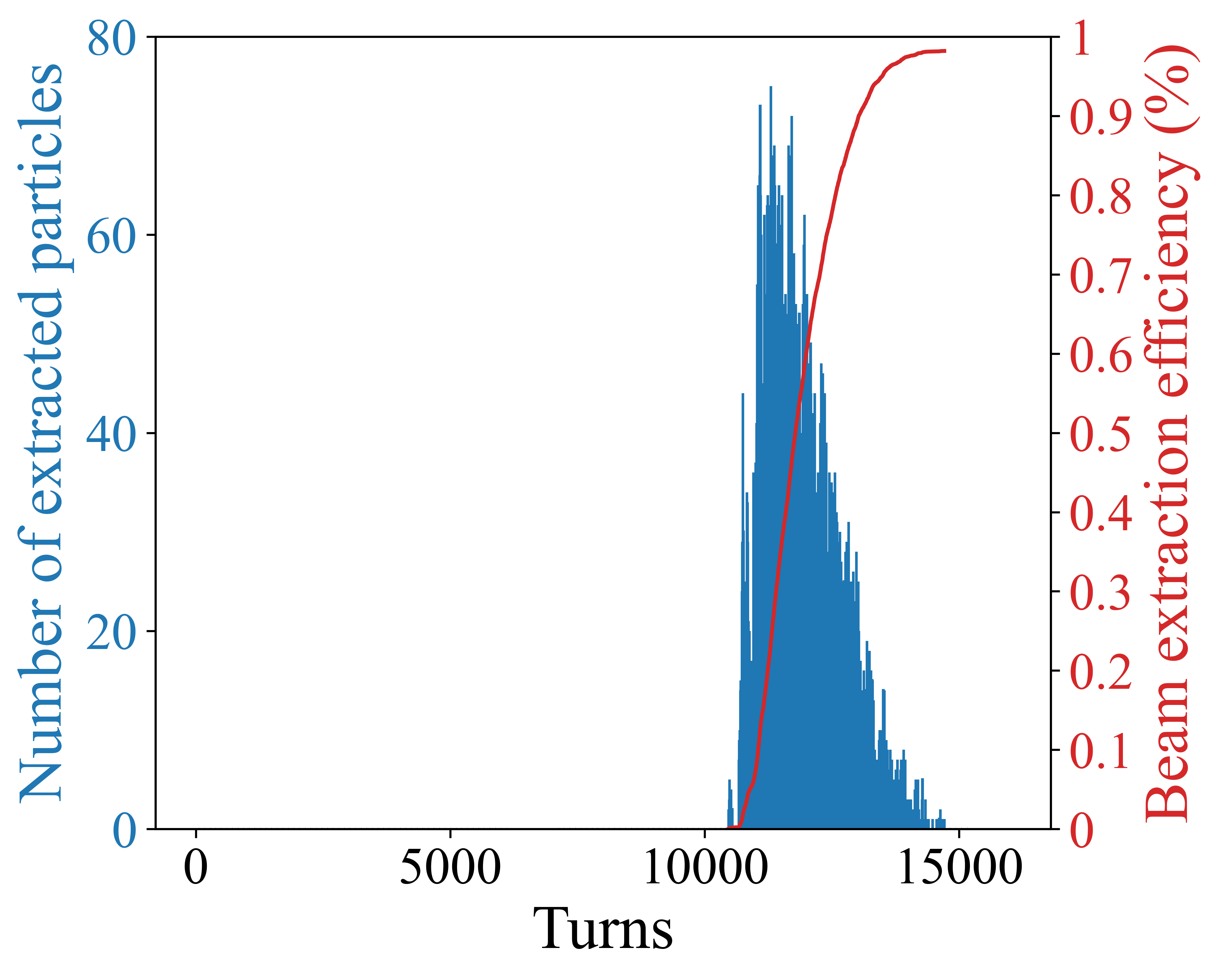}
  \end{minipage}
  \hspace{0.001\textwidth} 
  \begin{minipage}{0.23\textwidth}
    \centering
    \includegraphics[width=\linewidth]{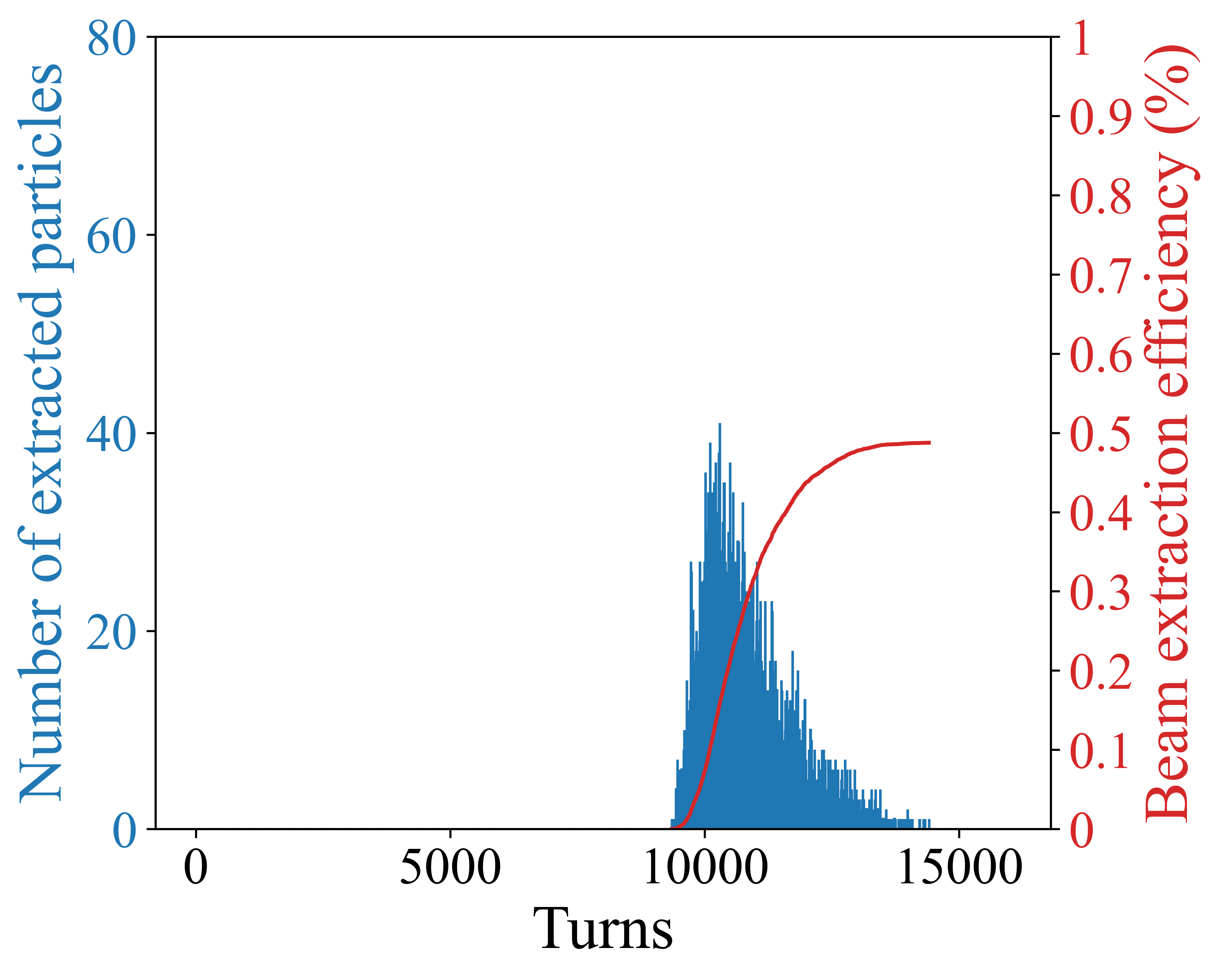}
  \end{minipage}

  \caption{\label{fig:emitt}Comparison of the simulated beam distribution in phase space of the 9400th turn and the spill structure for beam initial emittance of 14 $\mu m$ (left) and 107 $\mu m$ (right).}
\end{figure}

With an initial emittance of $14 ~\mu m$, during the simulation process, all particle's amplitude are simultaneously excited. The beam forms a distinct "ring"-shaped pattern in the phase space, as shown in the upper left figure in Fig.~\ref{fig:emitt}, indicating that the entire beam achieves autoresonance. Beam extraction begins around the 10500th turn, and by the 15000th turn, the extraction efficiency reaches $98.2\%$.

With an initial emittance of $107 ~\mu m$, after a certain period, the oscillation amplitudes of a part of the particles stop increasing, and the particles in the "blue ring" will be extracted subsequently. There are also literature sources that present phase space distributions similar to our simulations \cite{cortes_garcia_interpretation_2024}. What happens during one frequency sweep period in conventional slow extraction is quite similar to this situation, where a part of particles are extracted with each sweep period. As shown in the upper right figure in Fig.~\ref{fig:emitt}, these unextracted particles are distinctly marked in red, representing those with larger initial emittances. It is evident that particles with larger initial emittances find it more challenging to meet the autoresonance conditions, making it difficult for them to achieve larger amplitudes. By the 15000th turn, the beam extraction efficiency is only $48.8\%$. To improve the extraction efficiency, one viable solution is to deepen the potential well by either increasing the excitation amplitude or adjusting the detuning parameter, as per the methods outlined in Eq.~(\ref{eq_final1}). This approach makes it possible for particles that are challenging to excite to larger amplitudes to resonate and be successfully extracted. Both initial emittance condition's spill structure exhibits a Rayleigh distribution, consistent with the beam spill structure observed in one of the periods during the conventional slow extraction process.

\begin{figure}[hbt]
  \centering 
  \includegraphics[width=8.4cm]{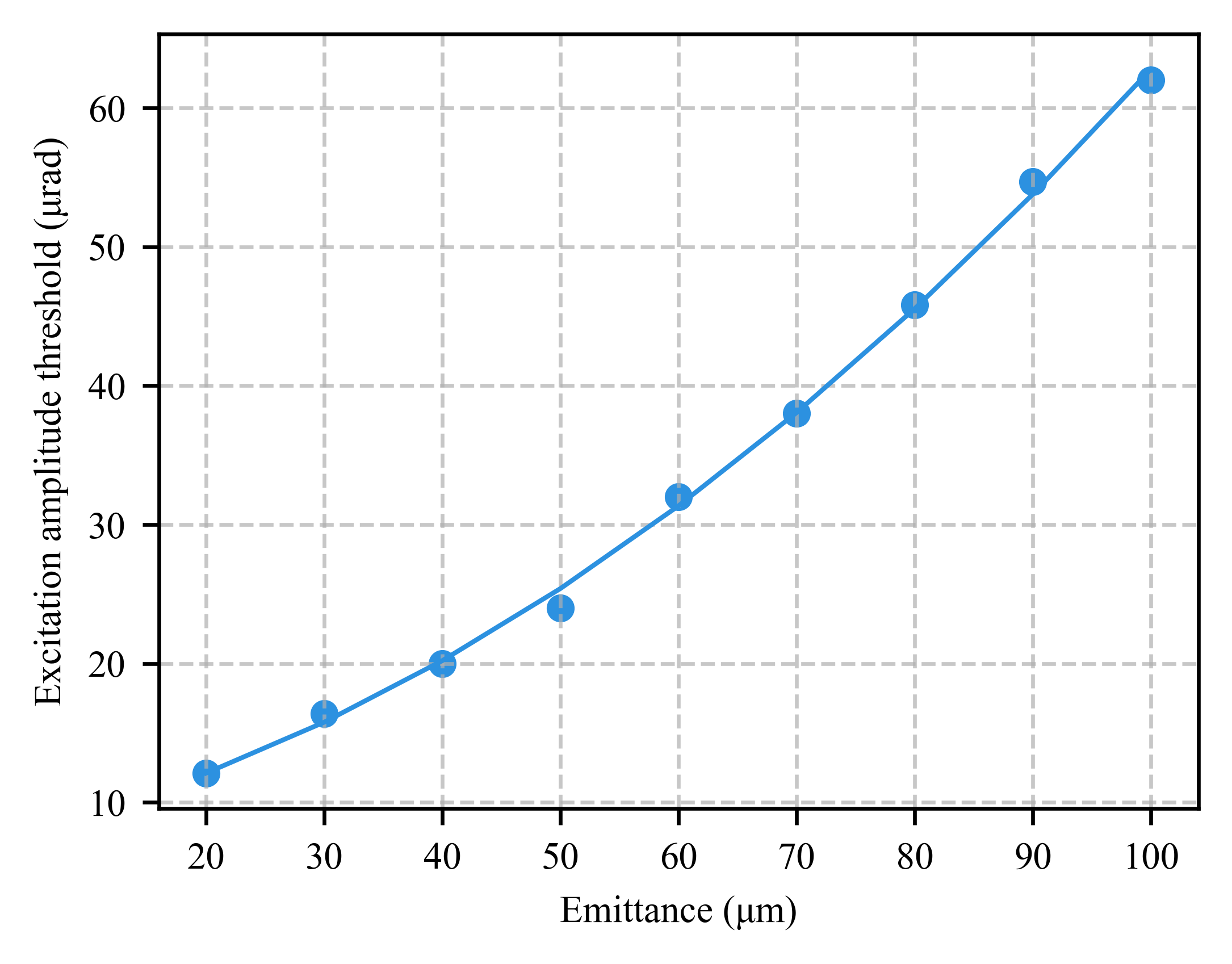} 
  \caption{\label{fig:emitt_and_threshold}Simulation result of the dependence of the excitation amplitude threshold on the beam emittance.}
\end{figure}

Fig.~\ref{fig:emitt_and_threshold} illustrates the dependence of the excitation amplitude threshold on beam emittance through simulation results. By simulating 10,000 particles, we aim to bring 99.7\% of them into autoresonance and record the minimum excitation amplitude threshold required with initial emittances ranging from $10 ~\mu m$ to $100 ~\mu m$. The results further validate the findings in Fig.~\ref{fig:emitt}, demonstrating that particles with larger initial emittance are more difficult to achieve autoresonance, thus requiring a higher excitation amplitude threshold.

\begin{figure}[hbt]
  \centering
  \begin{minipage}{0.23\textwidth}
    \centering
    \includegraphics[width=\linewidth]{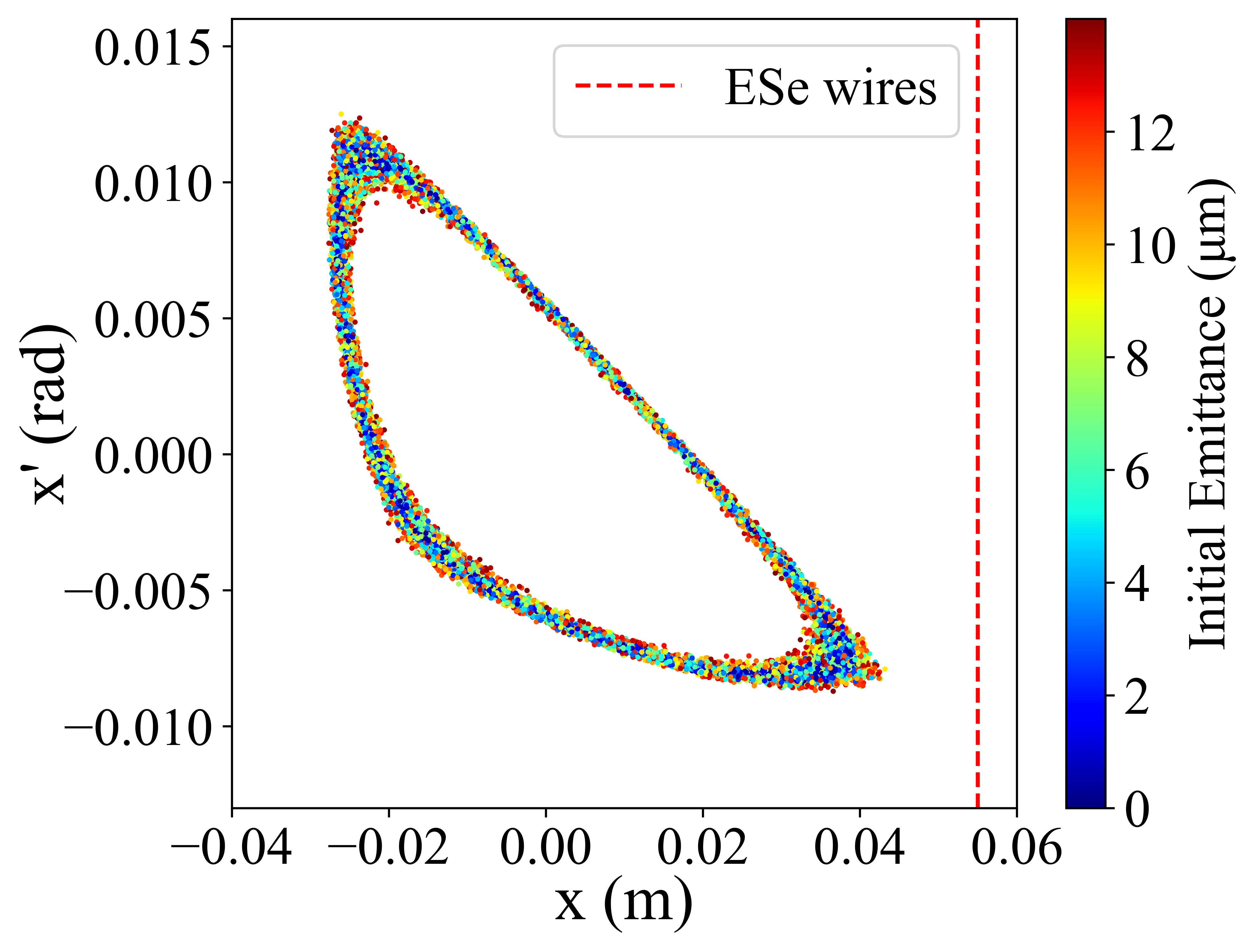}
  \end{minipage}
  \hspace{0.004\textwidth} 
  \begin{minipage}{0.23\textwidth}
    \centering
    \includegraphics[width=\linewidth]{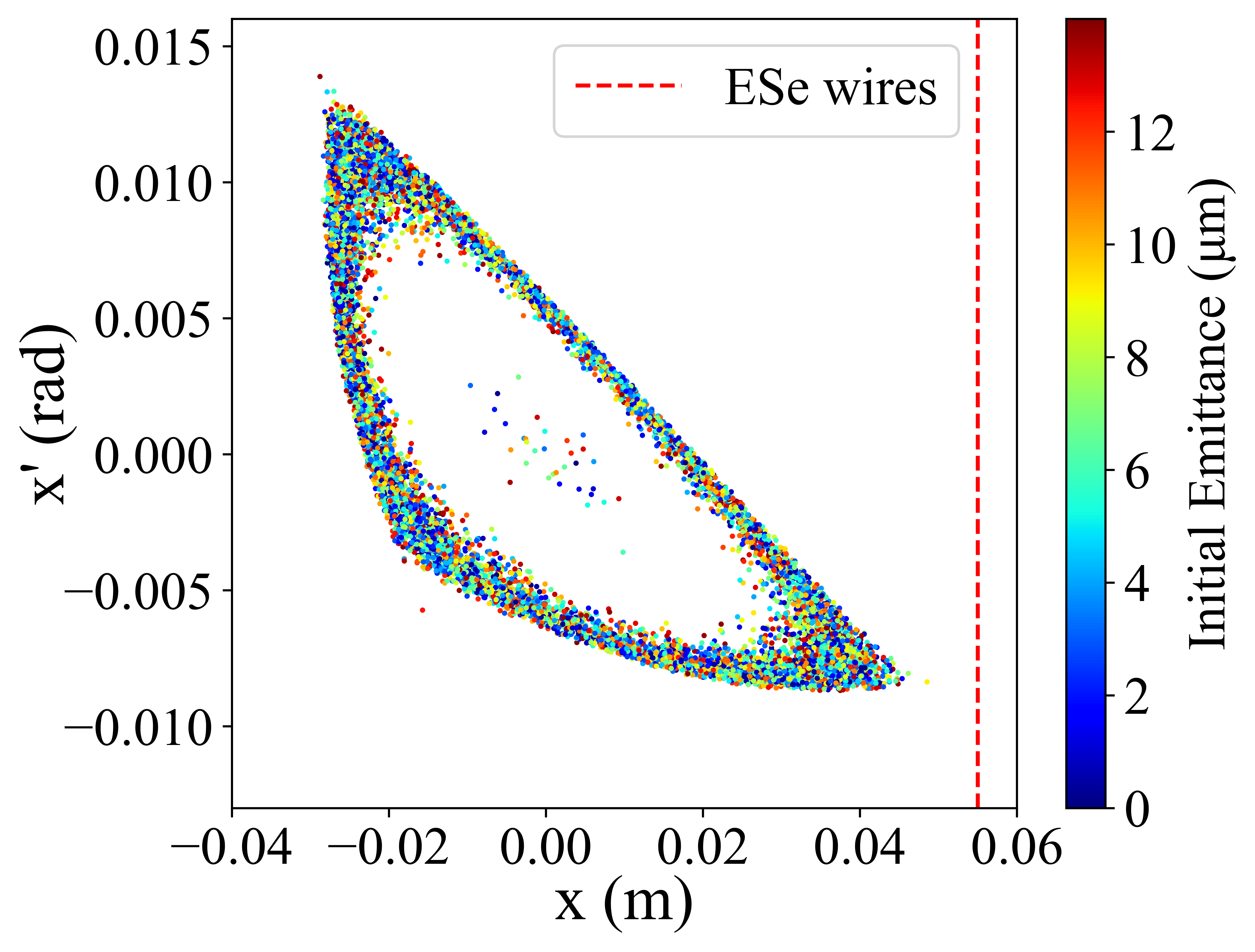}
  \end{minipage}

  \vspace{0.5em} 

  \begin{minipage}{0.23\textwidth}
    \centering
    \includegraphics[width=\linewidth]{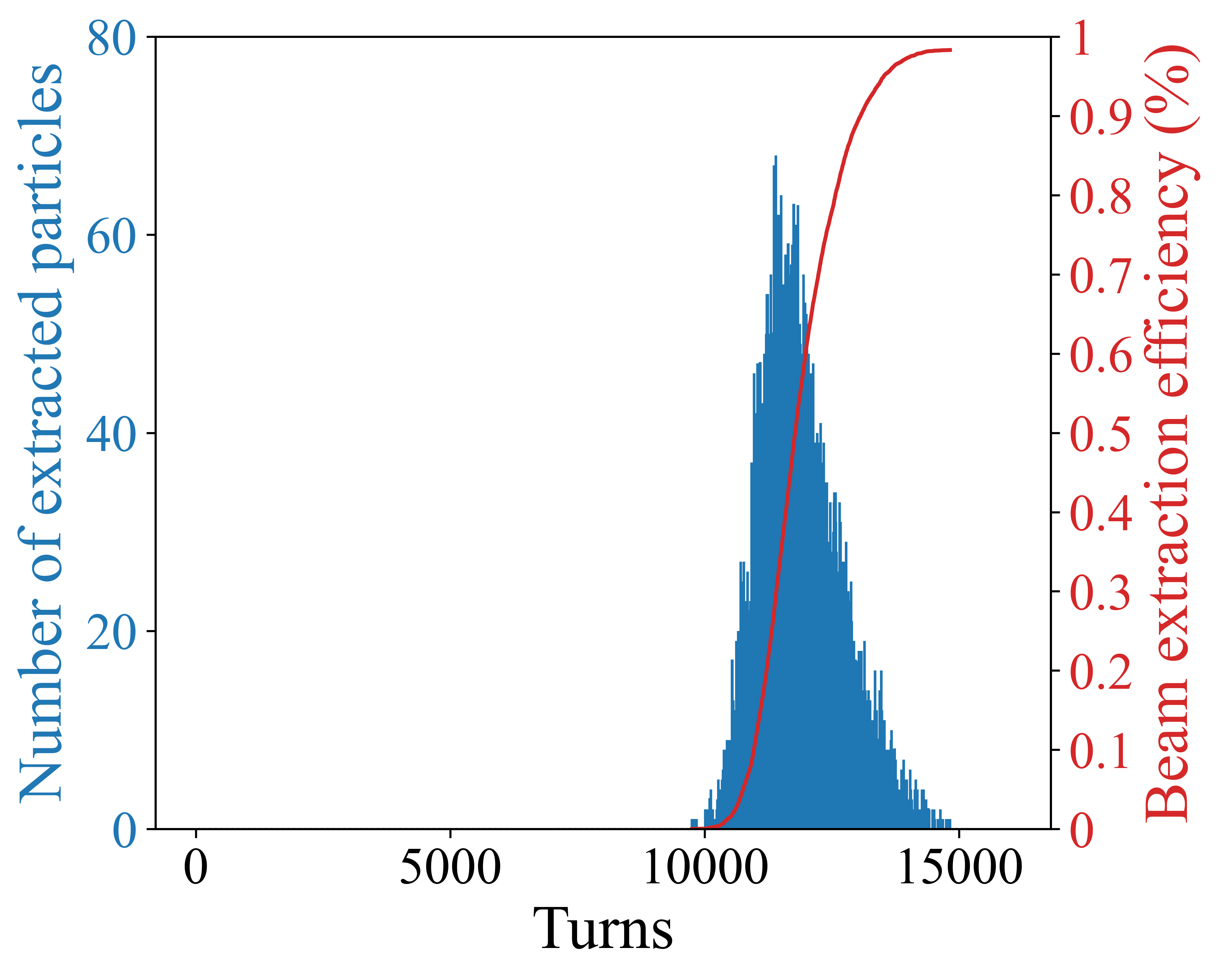}
  \end{minipage}
  \hspace{0.001\textwidth} 
  \begin{minipage}{0.23\textwidth}
    \centering
    \includegraphics[width=\linewidth]{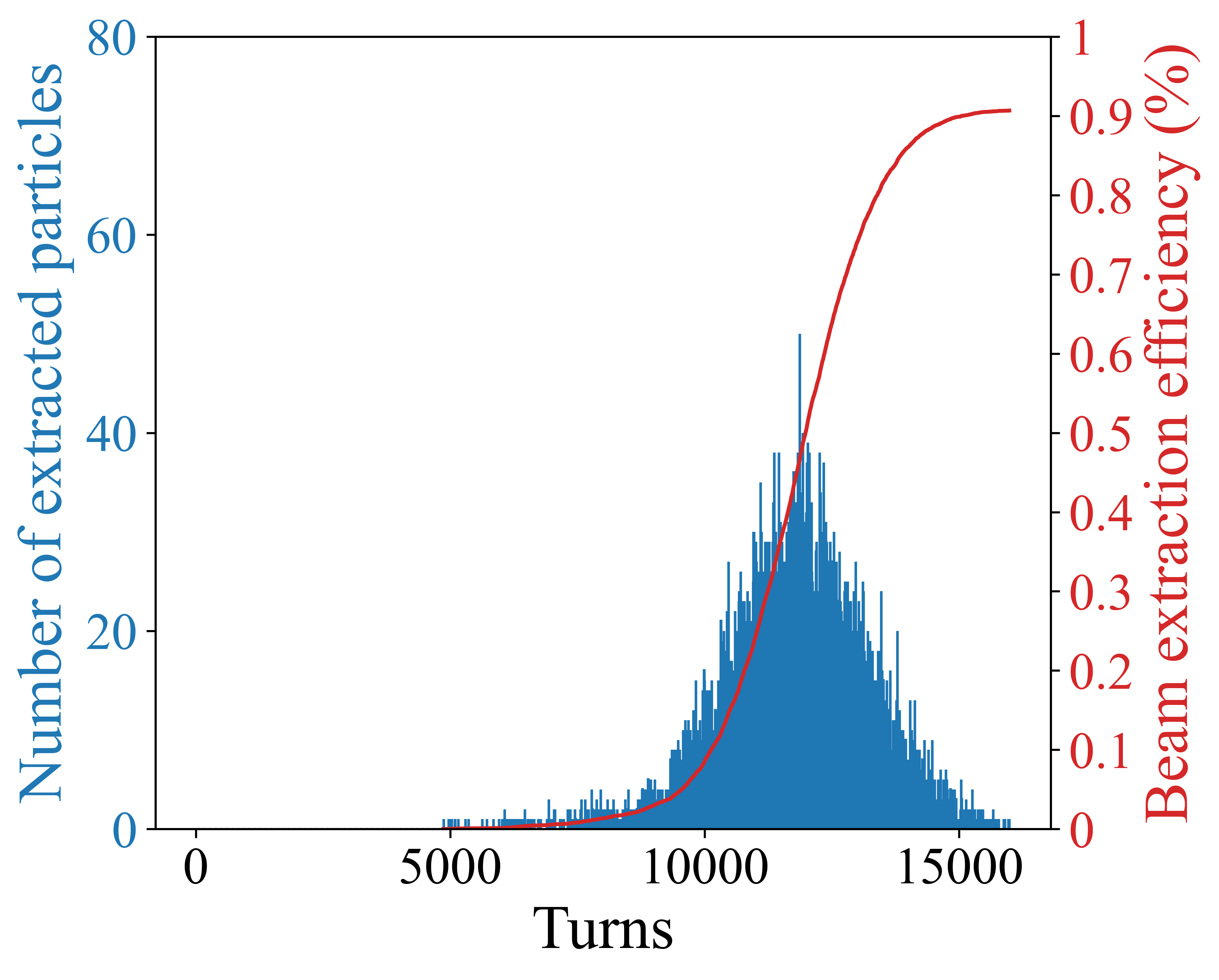}
  \end{minipage}

  \caption{\label{fig:momentum_spreads}Comparison of the simulated beam distribution in phase space and the extracted structure for beam momentum spreads of 0.1\% and 0.5\%.}
\end{figure}

In addition, simulations with momentum spreads of $0.1\%$ and $0.5\%$ are conducted as Fig.~\ref{fig:momentum_spreads}. Due to the existence of a 3.4m dispersion at the observation point, the "ring" formed by the beam with a larger momentum spread is wider. Consequently, this leads to a phenomenon where the extraction structure broadens the original Rayleigh distribution. The chromaticity within the ring is relatively small, which should have little impact on the tune. The figure shows that the momentum spread has little effect on the autoresonance, and as long as the frequency sweep can cover the corresponding range, a larger momentum spread can still cause entire beam autoresonance. The simulation results indicate that neither autoresonance excitation nor conventional RF-KO excitation will disrupt the particle motion during the third-order resonance slow extraction process.

\subsection{Autoresonance rapid extraction advantage}

Compared to conventional slow extraction, autoresonance rapid extraction exhibits several notable differences. Firstly, particles in autoresonance state are characterized by a phase locking of their oscillation frequency to the excitation frequency, allowing for strong excitation of particle amplitude with a smaller excitation amplitude or facilitating a reduction in extraction time to the millisecond range. Secondly, autoresonance rapid extraction requires strong detuning, requiring that the initial tune be far away from the third-order resonance line, which provides additional stability for the beam. Additionally, autoresonance rapid extraction eliminates the need for quadrupole magnets to modify the tuning, thereby simplifying the extraction process and potentially further reducing the extraction time. Thirdly, during the amplitude excitation process in autoresonance rapid extraction, the action in phase space, which corresponds to the particle's amplitude, is primarily governed by detuning. The detuning is solely determined by the parameters of the nonlinear magnets in phase space. Consequently, precise control over the particle's amplitude can be achieved by merely adjusting the nonlinear magnets and the driver.

\section{conclusion}
The autoresonance technique enables nonlinear oscillators to achieve large oscillation amplitude by applying a swept frequency excitation, where the beam remains phase-locked with the drive once the excitation amplitude exceeds a specific threshold. In this study, we derived the autoresonance threshold for an oscillator under the influence of sextupole and octupole magnetic fields, with single particle simulation showing good agreement with the theoretical formula. Benefiting from the strong detuning effect introduced by octupole, autoresonance can be efficiently driven by an excitation with small amplitude. Accordingly, we propose a rapid beam extraction method for synchrotrons and demonstrate its feasibility through theoretical analysis and simulation.

In the proposed rapid extraction, the entire beam is brought into autoresonance by aligning the excitation amplitude with the threshold conditions. As the oscillation  amplitude increases and approaches the third-order resonance, the separatrix dominated by sextupole field enables the particle to enter the extraction channel. This allows for beam extraction within one sweeping period on the order of millisecond. Simulations on SESRI 300 MeV synchrotron validate the theoretical framework and confirm the capability of millisecond slow extraction requirement for FLASH therapy using existing excitation techniques. Compared with the conventional RF-KO method, this innovative method requires only the addition of one octupole magnet, and the power requirement of the driver can be significantly reduced, mitigates potential risks associated with high-power drive. Concurrently, this method reduced need for tune adjustment also simplifies operational complexity and may further reduce the proportion of the extraction platform throughout the entire acceleration cycle.

Furthermore, this autoresonance based rapid extraction technique is applicable in various contexts, such as Rapid Cycling Synchrotrons (RCS). Our findings suggest that it offers a promising solution to the challenges of rapid beam extraction in synchrotrons, providing a more efficient and safer operational pathway for both medical applications and high-energy physics research. Future efforts would concentrate on the experimental implementation of this technique in operational synchrotron, as well as on further optimization of parameters for extraction efficiency and beam quality.

\section{Acknowledgment}

The authors would like to thank colleagues in the accelerator physics group at IMP for useful suggestions and discussions. This work is supported by the National Natural Science Foundation of China (Grant No. 12425501, No. 12275323, No. 12405177), the Hundred Talents Project of the Chinese Academy of Sciences, the Natural Science Foundation of Gansu Province, China (Grant No. 25JRRA459), and the Gansu Intellectual Property Plan (Grant No. 24ZSCQD002).


\bibliography{Application_of_autoresonance_in_rapid_beam_extraction_of_synchrotrons}

\end{document}